\documentclass[aps,prl,twocolumn,showpacs,superscriptaddress,groupedaddress]{revtex4}  
\usepackage{graphicx}  
\usepackage{dcolumn}   
\usepackage{bm}        
\usepackage{amssymb}   
\usepackage{tikz}
\usepackage{float}
\usepackage{xcolor}
\usepackage{esvect}
\usepackage{amsmath}
\usepackage{diagbox}
\usepackage{amssymb}
\usepackage{tikz}
\usetikzlibrary{calc}
\usetikzlibrary{chains}
\usetikzlibrary{scopes}
\usetikzlibrary{shapes.geometric}
\usetikzlibrary{trees}
\usetikzlibrary{topaths}
\usetikzlibrary{positioning}
\usetikzlibrary{patterns,decorations.pathreplacing}

\newcommand{\bP}{\mathbb P}

\usepackage[vcentermath]{youngtab}
\usepackage{ytableau}

\usepackage{amssymb}
\usepackage{tikz}
\usetikzlibrary{calc}
\usetikzlibrary{chains}
\usetikzlibrary{scopes}
\usetikzlibrary{shapes.geometric}
\usetikzlibrary{trees}
\usetikzlibrary{topaths}
\usetikzlibrary{positioning}
\usetikzlibrary{patterns,decorations.pathreplacing}

\newcommand{\Pabcd}{
 \left( 
 \begin{tikzpicture}
[baseline=1pt]
 \draw [fill] (0,0) circle(0.2ex);
 \draw [fill] (0.10,0) circle(0.2ex);
 \draw [fill] (0.20,0) circle(0.2ex);
 \draw [fill] (0.30,0) circle(0.2ex);
 \draw [fill] (0,0.25) circle(0.2ex);
 \draw [fill] (0.10,0.25) circle(0.2ex);
 \draw [fill] (0.20,0.25) circle(0.2ex);
 \draw [fill] (0.30,0.25) circle(0.2ex);
 \draw [thick] (0.00,0.25)  --(0.00,0.00);
 \draw [thick] (0.10,0.25)--(0.10,0.00);
 \draw [thick] (0.20,0.25)--(0.20,0.00);
 \draw [thick] (0.30,0.25)--(0.30,0.00);
 \end{tikzpicture}
\right)
}

\newcommand{\Pabdc}{ 
\left( 
 \begin{tikzpicture}
[baseline=1pt]
 \draw [fill] (0,0) circle(0.2ex);
 \draw [fill] (0.10,0) circle(0.2ex);
 \draw [fill] (0.20,0) circle(0.2ex);
 \draw [fill] (0.30,0) circle(0.2ex);
 \draw [fill] (0,0.25) circle(0.2ex);
 \draw [fill] (0.10,0.25) circle(0.2ex);
 \draw [fill] (0.20,0.25) circle(0.2ex);
 \draw [fill] (0.30,0.25) circle(0.2ex);
 \draw [thick] (0.00,0.25)  --(0.00,0.00);
 \draw [thick] (0.10,0.25)--(0.10,0.00);
 \draw [thick] (0.20,0.25)--(0.30,0.00);
 \draw [thick] (0.30,0.25)--(0.20,0.00);
 \end{tikzpicture}
\right)
 }

\newcommand{\Pacbd}{ 
\left( 
 \begin{tikzpicture}
[baseline=1pt]
 \draw [fill] (0,0) circle(0.2ex);
 \draw [fill] (0.10,0) circle(0.2ex);
 \draw [fill] (0.20,0) circle(0.2ex);
 \draw [fill] (0.30,0) circle(0.2ex);
 \draw [fill] (0,0.25) circle(0.2ex);
 \draw [fill] (0.10,0.25) circle(0.2ex);
 \draw [fill] (0.20,0.25) circle(0.2ex);
 \draw [fill] (0.30,0.25) circle(0.2ex);
 \draw [thick] (0.00,0.25)  --(0.00,0.00);
 \draw [thick] (0.10,0.25)--(0.20,0.00);
 \draw [thick] (0.20,0.25)--(0.10,0.00);
 \draw [thick] (0.30,0.25)--(0.30,0.00);
 \end{tikzpicture}
\right)
 }

\newcommand{\Pacdb}{ 
\left( 
 \begin{tikzpicture}
[baseline=1pt]
 \draw [fill] (0,0) circle(0.2ex);
 \draw [fill] (0.10,0) circle(0.2ex);
 \draw [fill] (0.20,0) circle(0.2ex);
 \draw [fill] (0.30,0) circle(0.2ex);
 \draw [fill] (0,0.25) circle(0.2ex);
 \draw [fill] (0.10,0.25) circle(0.2ex);
 \draw [fill] (0.20,0.25) circle(0.2ex);
 \draw [fill] (0.30,0.25) circle(0.2ex);
 \draw [thick] (0.00,0.25)  --(0.00,0.00);
 \draw [thick] (0.10,0.25)--(0.20,0.00);
 \draw [thick] (0.20,0.25)--(0.30,0.00);
 \draw [thick] (0.30,0.25)--(0.10,0.00);
 \end{tikzpicture}
\right)
 }

\newcommand{\Padbc}{ 
\left( 
 \begin{tikzpicture}
[baseline=1pt]
 \draw [fill] (0,0) circle(0.2ex);
 \draw [fill] (0.10,0) circle(0.2ex);
 \draw [fill] (0.20,0) circle(0.2ex);
 \draw [fill] (0.30,0) circle(0.2ex);
 \draw [fill] (0,0.25) circle(0.2ex);
 \draw [fill] (0.10,0.25) circle(0.2ex);
 \draw [fill] (0.20,0.25) circle(0.2ex);
 \draw [fill] (0.30,0.25) circle(0.2ex);
 \draw [thick] (0.00,0.25)  --(0.00,0.00);
 \draw [thick] (0.10,0.25)--(0.30,0.00);
 \draw [thick] (0.20,0.25)--(0.10,0.00);
 \draw [thick] (0.30,0.25)--(0.20,0.00);
 \end{tikzpicture}
\right)
 }

\newcommand{\Padcb}{ 
\left( 
 \begin{tikzpicture}
[baseline=1pt]
 \draw [fill] (0,0) circle(0.2ex);
 \draw [fill] (0.10,0) circle(0.2ex);
 \draw [fill] (0.20,0) circle(0.2ex);
 \draw [fill] (0.30,0) circle(0.2ex);
 \draw [fill] (0,0.25) circle(0.2ex);
 \draw [fill] (0.10,0.25) circle(0.2ex);
 \draw [fill] (0.20,0.25) circle(0.2ex);
 \draw [fill] (0.30,0.25) circle(0.2ex);
 \draw [thick] (0.00,0.25)  --(0.00,0.00);
 \draw [thick] (0.10,0.25)--(0.30,0.00);
 \draw [thick] (0.20,0.25)--(0.20,0.00);
 \draw [thick] (0.30,0.25)--(0.10,0.00);
 \end{tikzpicture}
\right)
 }

\newcommand{\Pbacd}{ 
\left( 
 \begin{tikzpicture}
[baseline=1pt]
 \draw [fill] (0,0) circle(0.2ex);
 \draw [fill] (0.10,0) circle(0.2ex);
 \draw [fill] (0.20,0) circle(0.2ex);
 \draw [fill] (0.30,0) circle(0.2ex);
 \draw [fill] (0,0.25) circle(0.2ex);
 \draw [fill] (0.10,0.25) circle(0.2ex);
 \draw [fill] (0.20,0.25) circle(0.2ex);
 \draw [fill] (0.30,0.25) circle(0.2ex);
 \draw [thick] (0.00,0.25)  --(0.10,0.00);
 \draw [thick] (0.10,0.25)--(0.00,0.00);
 \draw [thick] (0.20,0.25)--(0.20,0.00);
 \draw [thick] (0.30,0.25)--(0.30,0.00);
 \end{tikzpicture}
\right)
 }

\newcommand{\Pbadc}{ 
\left( 
 \begin{tikzpicture}
[baseline=1pt]
 \draw [fill] (0,0) circle(0.2ex);
 \draw [fill] (0.10,0) circle(0.2ex);
 \draw [fill] (0.20,0) circle(0.2ex);
 \draw [fill] (0.30,0) circle(0.2ex);
 \draw [fill] (0,0.25) circle(0.2ex);
 \draw [fill] (0.10,0.25) circle(0.2ex);
 \draw [fill] (0.20,0.25) circle(0.2ex);
 \draw [fill] (0.30,0.25) circle(0.2ex);
 \draw [thick] (0.00,0.25)  --(0.10,0.00);
 \draw [thick] (0.10,0.25)--(0.00,0.00);
 \draw [thick] (0.20,0.25)--(0.30,0.00);
 \draw [thick] (0.30,0.25)--(0.20,0.00);
 \end{tikzpicture}
\right)
 }

\newcommand{\Pbcad}{ 
\left( 
 \begin{tikzpicture}
[baseline=1pt]
 \draw [fill] (0,0) circle(0.2ex);
 \draw [fill] (0.10,0) circle(0.2ex);
 \draw [fill] (0.20,0) circle(0.2ex);
 \draw [fill] (0.30,0) circle(0.2ex);
 \draw [fill] (0,0.25) circle(0.2ex);
 \draw [fill] (0.10,0.25) circle(0.2ex);
 \draw [fill] (0.20,0.25) circle(0.2ex);
 \draw [fill] (0.30,0.25) circle(0.2ex);
 \draw [thick] (0.00,0.25)  --(0.10,0.00);
 \draw [thick] (0.10,0.25)--(0.20,0.00);
 \draw [thick] (0.20,0.25)--(0.00,0.00);
 \draw [thick] (0.30,0.25)--(0.30,0.00);
 \end{tikzpicture}
\right)
 }

\newcommand{\Pbcda}{ 
\left( 
 \begin{tikzpicture}
[baseline=1pt]
 \draw [fill] (0,0) circle(0.2ex);
 \draw [fill] (0.10,0) circle(0.2ex);
 \draw [fill] (0.20,0) circle(0.2ex);
 \draw [fill] (0.30,0) circle(0.2ex);
 \draw [fill] (0,0.25) circle(0.2ex);
 \draw [fill] (0.10,0.25) circle(0.2ex);
 \draw [fill] (0.20,0.25) circle(0.2ex);
 \draw [fill] (0.30,0.25) circle(0.2ex);
 \draw [thick] (0.00,0.25)  --(0.10,0.00);
 \draw [thick] (0.10,0.25)--(0.20,0.00);
 \draw [thick] (0.20,0.25)--(0.30,0.00);
 \draw [thick] (0.30,0.25)--(0.00,0.00);
 \end{tikzpicture}
\right)
 }

\newcommand{\Pbdac}{ 
\left( 
 \begin{tikzpicture}
[baseline=1pt]
 \draw [fill] (0,0) circle(0.2ex);
 \draw [fill] (0.10,0) circle(0.2ex);
 \draw [fill] (0.20,0) circle(0.2ex);
 \draw [fill] (0.30,0) circle(0.2ex);
 \draw [fill] (0,0.25) circle(0.2ex);
 \draw [fill] (0.10,0.25) circle(0.2ex);
 \draw [fill] (0.20,0.25) circle(0.2ex);
 \draw [fill] (0.30,0.25) circle(0.2ex);
 \draw [thick] (0.00,0.25)  --(0.10,0.00);
 \draw [thick] (0.10,0.25)--(0.30,0.00);
 \draw [thick] (0.20,0.25)--(0.00,0.00);
 \draw [thick] (0.30,0.25)--(0.20,0.00);
 \end{tikzpicture}
\right)
 }

\newcommand{\Pbdca}{ 
\left( 
 \begin{tikzpicture}
[baseline=1pt]
 \draw [fill] (0,0) circle(0.2ex);
 \draw [fill] (0.10,0) circle(0.2ex);
 \draw [fill] (0.20,0) circle(0.2ex);
 \draw [fill] (0.30,0) circle(0.2ex);
 \draw [fill] (0,0.25) circle(0.2ex);
 \draw [fill] (0.10,0.25) circle(0.2ex);
 \draw [fill] (0.20,0.25) circle(0.2ex);
 \draw [fill] (0.30,0.25) circle(0.2ex);
 \draw [thick] (0.00,0.25)  --(0.10,0.00);
 \draw [thick] (0.10,0.25)--(0.30,0.00);
 \draw [thick] (0.20,0.25)--(0.20,0.00);
 \draw [thick] (0.30,0.25)--(0.00,0.00);
 \end{tikzpicture}
\right)
 }

\newcommand{\Pcabd}{ 
\left( 
 \begin{tikzpicture}
[baseline=1pt]
 \draw [fill] (0,0) circle(0.2ex);
 \draw [fill] (0.10,0) circle(0.2ex);
 \draw [fill] (0.20,0) circle(0.2ex);
 \draw [fill] (0.30,0) circle(0.2ex);
 \draw [fill] (0,0.25) circle(0.2ex);
 \draw [fill] (0.10,0.25) circle(0.2ex);
 \draw [fill] (0.20,0.25) circle(0.2ex);
 \draw [fill] (0.30,0.25) circle(0.2ex);
 \draw [thick] (0.00,0.25)  --(0.20,0.00);
 \draw [thick] (0.10,0.25)--(0.00,0.00);
 \draw [thick] (0.20,0.25)--(0.10,0.00);
 \draw [thick] (0.30,0.25)--(0.30,0.00);
 \end{tikzpicture}
\right)
 }
\newcommand{\Pcadb}{ 
\left( 
 \begin{tikzpicture}
[baseline=1pt]
 \draw [fill] (0,0) circle(0.2ex);
 \draw [fill] (0.10,0) circle(0.2ex);
 \draw [fill] (0.20,0) circle(0.2ex);
 \draw [fill] (0.30,0) circle(0.2ex);
 \draw [fill] (0,0.25) circle(0.2ex);
 \draw [fill] (0.10,0.25) circle(0.2ex);
 \draw [fill] (0.20,0.25) circle(0.2ex);
 \draw [fill] (0.30,0.25) circle(0.2ex);
 \draw [thick] (0.00,0.25)  --(0.20,0.00);
 \draw [thick] (0.10,0.25)--(0.00,0.00);
 \draw [thick] (0.20,0.25)--(0.30,0.00);
 \draw [thick] (0.30,0.25)--(0.10,0.00);
 \end{tikzpicture}
\right)
 }
\newcommand{\Pcbad}{ 
\left( 
 \begin{tikzpicture}
[baseline=1pt]
 \draw [fill] (0,0) circle(0.2ex);
 \draw [fill] (0.10,0) circle(0.2ex);
 \draw [fill] (0.20,0) circle(0.2ex);
 \draw [fill] (0.30,0) circle(0.2ex);
 \draw [fill] (0,0.25) circle(0.2ex);
 \draw [fill] (0.10,0.25) circle(0.2ex);
 \draw [fill] (0.20,0.25) circle(0.2ex);
 \draw [fill] (0.30,0.25) circle(0.2ex);
 \draw [thick] (0.00,0.25)  --(0.20,0.00);
 \draw [thick] (0.10,0.25)--(0.10,0.00);
 \draw [thick] (0.20,0.25)--(0.00,0.00);
 \draw [thick] (0.30,0.25)--(0.30,0.00);
 \end{tikzpicture}
\right)
 }
\newcommand{\Pcbda}{ 
\left( 
 \begin{tikzpicture}
[baseline=1pt]
 \draw [fill] (0,0) circle(0.2ex);
 \draw [fill] (0.10,0) circle(0.2ex);
 \draw [fill] (0.20,0) circle(0.2ex);
 \draw [fill] (0.30,0) circle(0.2ex);
 \draw [fill] (0,0.25) circle(0.2ex);
 \draw [fill] (0.10,0.25) circle(0.2ex);
 \draw [fill] (0.20,0.25) circle(0.2ex);
 \draw [fill] (0.30,0.25) circle(0.2ex);
 \draw [thick] (0.00,0.25)  --(0.20,0.00);
 \draw [thick] (0.10,0.25)--(0.10,0.00);
 \draw [thick] (0.20,0.25)--(0.30,0.00);
 \draw [thick] (0.30,0.25)--(0.00,0.00);
 \end{tikzpicture}
\right)
 }
\newcommand{\Pcdab}{ 
\left( 
 \begin{tikzpicture}
[baseline=1pt]
 \draw [fill] (0,0) circle(0.2ex);
 \draw [fill] (0.10,0) circle(0.2ex);
 \draw [fill] (0.20,0) circle(0.2ex);
 \draw [fill] (0.30,0) circle(0.2ex);
 \draw [fill] (0,0.25) circle(0.2ex);
 \draw [fill] (0.10,0.25) circle(0.2ex);
 \draw [fill] (0.20,0.25) circle(0.2ex);
 \draw [fill] (0.30,0.25) circle(0.2ex);
 \draw [thick] (0.00,0.25)  --(0.20,0.00);
 \draw [thick] (0.10,0.25)--(0.30,0.00);
 \draw [thick] (0.20,0.25)--(0.00,0.00);
 \draw [thick] (0.30,0.25)--(0.10,0.00);
 \end{tikzpicture}
\right)
 }
\newcommand{\Pcdba}{ 
\left( 
 \begin{tikzpicture}
[baseline=1pt]
 \draw [fill] (0,0) circle(0.2ex);
 \draw [fill] (0.10,0) circle(0.2ex);
 \draw [fill] (0.20,0) circle(0.2ex);
 \draw [fill] (0.30,0) circle(0.2ex);
 \draw [fill] (0,0.25) circle(0.2ex);
 \draw [fill] (0.10,0.25) circle(0.2ex);
 \draw [fill] (0.20,0.25) circle(0.2ex);
 \draw [fill] (0.30,0.25) circle(0.2ex);
 \draw [thick] (0.00,0.25)  --(0.20,0.00);
 \draw [thick] (0.10,0.25)--(0.30,0.00);
 \draw [thick] (0.20,0.25)--(0.10,0.00);
 \draw [thick] (0.30,0.25)--(0.00,0.00);
 \end{tikzpicture}
\right)
 }

\newcommand{\Pdabc}{ 
\left( 
 \begin{tikzpicture}
[baseline=1pt]
 \draw [fill] (0,0) circle(0.2ex);
 \draw [fill] (0.10,0) circle(0.2ex);
 \draw [fill] (0.20,0) circle(0.2ex);
 \draw [fill] (0.30,0) circle(0.2ex);
 \draw [fill] (0,0.25) circle(0.2ex);
 \draw [fill] (0.10,0.25) circle(0.2ex);
 \draw [fill] (0.20,0.25) circle(0.2ex);
 \draw [fill] (0.30,0.25) circle(0.2ex);
 \draw [thick] (0.00,0.25)  --(0.30,0.00);
 \draw [thick] (0.10,0.25)--(0.00,0.00);
 \draw [thick] (0.20,0.25)--(0.10,0.00);
 \draw [thick] (0.30,0.25)--(0.20,0.00);
 \end{tikzpicture}
\right)
 }
\newcommand{\Pdacb}{ 
\left( 
 \begin{tikzpicture}
[baseline=1pt]
 \draw [fill] (0,0) circle(0.2ex);
 \draw [fill] (0.10,0) circle(0.2ex);
 \draw [fill] (0.20,0) circle(0.2ex);
 \draw [fill] (0.30,0) circle(0.2ex);
 \draw [fill] (0,0.25) circle(0.2ex);
 \draw [fill] (0.10,0.25) circle(0.2ex);
 \draw [fill] (0.20,0.25) circle(0.2ex);
 \draw [fill] (0.30,0.25) circle(0.2ex);
 \draw [thick] (0.00,0.25)  --(0.30,0.00);
 \draw [thick] (0.10,0.25)--(0.00,0.00);
 \draw [thick] (0.20,0.25)--(0.20,0.00);
 \draw [thick] (0.30,0.25)--(0.10,0.00);
 \end{tikzpicture}
\right)
 }
\newcommand{\Pdbac}{ 
\left( 
 \begin{tikzpicture}
[baseline=1pt]
 \draw [fill] (0,0) circle(0.2ex);
 \draw [fill] (0.10,0) circle(0.2ex);
 \draw [fill] (0.20,0) circle(0.2ex);
 \draw [fill] (0.30,0) circle(0.2ex);
 \draw [fill] (0,0.25) circle(0.2ex);
 \draw [fill] (0.10,0.25) circle(0.2ex);
 \draw [fill] (0.20,0.25) circle(0.2ex);
 \draw [fill] (0.30,0.25) circle(0.2ex);
 \draw [thick] (0.00,0.25)  --(0.30,0.00);
 \draw [thick] (0.10,0.25)--(0.10,0.00);
 \draw [thick] (0.20,0.25)--(0.00,0.00);
 \draw [thick] (0.30,0.25)--(0.20,0.00);
 \end{tikzpicture}
\right)
 }
\newcommand{\Pdbca}{ 
\left( 
 \begin{tikzpicture}
[baseline=1pt]
 \draw [fill] (0,0) circle(0.2ex);
 \draw [fill] (0.10,0) circle(0.2ex);
 \draw [fill] (0.20,0) circle(0.2ex);
 \draw [fill] (0.30,0) circle(0.2ex);
 \draw [fill] (0,0.25) circle(0.2ex);
 \draw [fill] (0.10,0.25) circle(0.2ex);
 \draw [fill] (0.20,0.25) circle(0.2ex);
 \draw [fill] (0.30,0.25) circle(0.2ex);
 \draw [thick] (0.00,0.25)  --(0.30,0.00);
 \draw [thick] (0.10,0.25)--(0.10,0.00);
 \draw [thick] (0.20,0.25)--(0.20,0.00);
 \draw [thick] (0.30,0.25)--(0.00,0.00);
 \end{tikzpicture}
\right)
 }
\newcommand{\Pdcab}{ 
\left( 
 \begin{tikzpicture}
[baseline=1pt]
 \draw [fill] (0,0) circle(0.2ex);
 \draw [fill] (0.10,0) circle(0.2ex);
 \draw [fill] (0.20,0) circle(0.2ex);
 \draw [fill] (0.30,0) circle(0.2ex);
 \draw [fill] (0,0.25) circle(0.2ex);
 \draw [fill] (0.10,0.25) circle(0.2ex);
 \draw [fill] (0.20,0.25) circle(0.2ex);
 \draw [fill] (0.30,0.25) circle(0.2ex);
 \draw [thick] (0.00,0.25)  --(0.30,0.00);
 \draw [thick] (0.10,0.25)--(0.20,0.00);
 \draw [thick] (0.20,0.25)--(0.00,0.00);
 \draw [thick] (0.30,0.25)--(0.10,0.00);
 \end{tikzpicture}
\right)
 }
\newcommand{\Pdcba}{ 
\left( 
 \begin{tikzpicture}
[baseline=1pt]
 \draw [fill] (0,0) circle(0.2ex);
 \draw [fill] (0.10,0) circle(0.2ex);
 \draw [fill] (0.20,0) circle(0.2ex);
 \draw [fill] (0.30,0) circle(0.2ex);
 \draw [fill] (0,0.25) circle(0.2ex);
 \draw [fill] (0.10,0.25) circle(0.2ex);
 \draw [fill] (0.20,0.25) circle(0.2ex);
 \draw [fill] (0.30,0.25) circle(0.2ex);
 \draw [thick] (0.00,0.25)  --(0.30,0.00);
 \draw [thick] (0.10,0.25)--(0.20,0.00);
 \draw [thick] (0.20,0.25)--(0.10,0.00);
 \draw [thick] (0.30,0.25)--(0.00,0.00);
 \end{tikzpicture}
\right)
 }

\def\bbp{{\mathbb P}}
\def\bfr{{\mathbf r}}

\newcommand{\np}{\;\;}
\newcommand*{\rom}[1]{\expandafter \romannumeral #1}

\newcommand{\PPzero}{
\left(
\begin{tikzpicture}[baseline=1pt]
 \draw [fill] (0,0) circle(0.2ex);
 \draw [fill] (0.1,0) circle(0.2ex);
 \draw [fill] (0,0.25) circle(0.2ex);
 \draw [fill] (0.1,0.25) circle(0.2ex);
\end{tikzpicture}
\right)}

\newcommand{\PPca}{
\left(
\begin{tikzpicture}[baseline=1pt]
 \draw [fill] (0,0) circle(0.2ex);
 \draw [fill] (0.1,0) circle(0.2ex);
 \draw [fill] (0,0.25) circle(0.2ex);
 \draw [fill] (0.1,0.25) circle(0.2ex);
 \draw [thick] (0,0)--(0,0.25);
\end{tikzpicture}
\right)}

\newcommand{\PPcb}{
\left(
\begin{tikzpicture}[baseline=1pt]
 \draw [fill] (0,0) circle(0.2ex);
 \draw [fill] (0.1,0) circle(0.2ex);
 \draw [fill] (0,0.25) circle(0.2ex);
 \draw [fill] (0.1,0.25) circle(0.2ex);
 \draw [thick] (0.1,0)--(0.1,0.25);
\end{tikzpicture}
\right)}

\newcommand{\PPcc}{
\left(
\begin{tikzpicture}[baseline=1pt]
 \draw [fill] (0,0) circle(0.2ex);
 \draw [fill] (0.1,0) circle(0.2ex);
 \draw [fill] (0,0.25) circle(0.2ex);
 \draw [fill] (0.1,0.25) circle(0.2ex);
 \draw [thick] (0,0)--(0.1,0.25);
\end{tikzpicture}
\right)}

\newcommand{\PPcd}{
\left(
\begin{tikzpicture}[baseline=1pt]
 \draw [fill] (0,0) circle(0.2ex);
 \draw [fill] (0.1,0) circle(0.2ex);
 \draw [fill] (0,0.25) circle(0.2ex);
 \draw [fill] (0.1,0.25) circle(0.2ex);
 \draw [thick] (0,0.25)--(0.1,0);
\end{tikzpicture}
\right)}

\newcommand{\PPa}{
\left(
\begin{tikzpicture}[baseline=1pt]
 \draw [fill] (0,0) circle(0.20ex);
 \draw [fill] (0.10,0) circle(0.20ex);
 \draw [fill] (0,0.25) circle(0.20ex);
 \draw [fill] (0.10,0.25) circle(0.20ex);
 \draw [thick] (0,0)--(0,0.25);
 \draw [thick] (0.10,0)--(0.10,0.25);
\end{tikzpicture}
\right)}

\newcommand{\PPb}{
\left(
\begin{tikzpicture}[baseline=1pt]
 \draw [fill] (0,0) circle(0.2ex);
 \draw [fill] (0.1,0) circle(0.2ex);
 \draw [fill] (0,0.25) circle(0.2ex);
 \draw [fill] (0.1,0.25) circle(0.2ex);
 \draw[thick] (0,0)--(0.1,0.25);
 \draw[thick] (0.1,0)--(0,0.25);
\end{tikzpicture}
\right)}

\newcommand{\PPPa}{
\left(
\begin{tikzpicture}[baseline=1pt]
 \draw [fill] (0,0) circle(0.2ex);
 \draw [fill] (0.1,0) circle(0.2ex);
 \draw [fill] (0.2,0) circle(0.2ex);
 \draw [fill] (0,0.25) circle(0.2ex);
 \draw [fill] (0.1,0.25) circle(0.2ex);
 \draw [fill] (0.2,0.25) circle(0.2ex);
  \draw [thick] (0,0)--(0,0.25);
  \draw [thick] (0.1,0)--(0.1,0.25);
  \draw [thick] (0.2,0)--(0.2,0.25);
\end{tikzpicture}
\right)}

\newcommand{\PPPb}{
\left(
\begin{tikzpicture}[baseline=1pt]
 \draw [fill] (0,0) circle(0.2ex);
 \draw [fill] (0.1,0) circle(0.2ex);
 \draw [fill] (0.2,0) circle(0.2ex);
 \draw [fill] (0,0.25) circle(0.2ex);
 \draw [fill] (0.1,0.25) circle(0.2ex);
 \draw [fill] (0.2,0.25) circle(0.2ex);
  \draw [thick] (0,0)--(0,0.25);
  \draw [thick] (0.1,0)--(0.2,0.25);
  \draw [thick] (0.2,0)--(0.1,0.25);
\end{tikzpicture}
\right)}

\newcommand{\PPPc}{
\left(
\begin{tikzpicture}[baseline=1pt]
 \draw [fill] (0,0) circle(0.2ex);
 \draw [fill] (0.1,0) circle(0.2ex);
 \draw [fill] (0.2,0) circle(0.2ex);
 \draw [fill] (0,0.25) circle(0.2ex);
 \draw [fill] (0.1,0.25) circle(0.2ex);
 \draw [fill] (0.2,0.25) circle(0.2ex);
  \draw [thick] (0,0)--(0.1,0.25);
  \draw [thick] (0.1,0)--(0.0,0.25);
  \draw [thick] (0.2,0)--(0.2,0.25);
\end{tikzpicture}
\right)}

\newcommand{\PPPd}{
\left(
\begin{tikzpicture}[baseline=1pt]
 \draw [fill] (0,0) circle(0.2ex);
 \draw [fill] (0.1,0) circle(0.2ex);
 \draw [fill] (0.2,0) circle(0.2ex);
 \draw [fill] (0,0.25) circle(0.2ex);
 \draw [fill] (0.1,0.25) circle(0.2ex);
 \draw [fill] (0.2,0.25) circle(0.2ex);
  \draw [thick] (0,0)--(0.1,0.25);
  \draw [thick] (0.1,0)--(0.2,0.25);
  \draw [thick] (0.2,0)--(0.0,0.25);
\end{tikzpicture}
\right)}

\newcommand{\PPPe}{
\left(
\begin{tikzpicture}[baseline=1pt]
 \draw [fill] (0,0) circle(0.2ex);
 \draw [fill] (0.1,0) circle(0.2ex);
 \draw [fill] (0.2,0) circle(0.2ex);
 \draw [fill] (0,0.25) circle(0.2ex);
 \draw [fill] (0.1,0.25) circle(0.2ex);
 \draw [fill] (0.2,0.25) circle(0.2ex);
  \draw [thick] (0,0)--(0.2,0.25);
  \draw [thick] (0.1,0)--(0.0,0.25);
  \draw [thick] (0.2,0)--(0.1,0.25);
\end{tikzpicture}
\right)}

\newcommand{\PPPf}{
\left(
\begin{tikzpicture}[baseline=1pt]
 \draw [fill] (0,0) circle(0.2ex);
 \draw [fill] (0.1,0) circle(0.2ex);
 \draw [fill] (0.2,0) circle(0.2ex);
 \draw [fill] (0,0.25) circle(0.2ex);
 \draw [fill] (0.1,0.25) circle(0.2ex);
 \draw [fill] (0.2,0.25) circle(0.2ex);
  \draw [thick] (0,0)--(0.2,0.25);
  \draw [thick] (0.1,0)--(0.1,0.25);
  \draw [thick] (0.2,0)--(0.0,0.25);
\end{tikzpicture}
\right)}

\begin{document}

\title{Observation of non-scalar and logarithmic correlations in 2D and 3D percolation}
\author{Xiaojun Tan}
\affiliation{Hefei National Laboratory for Physical Sciences at Microscale and Department of Modern Physics, University of Science and Technology of China, Hefei, Anhui 230026, China}
\affiliation{CAS Center for Excellence and Synergetic Innovation Center in Quantum Information and Quantum Physics, University of Science and Technology of China, Hefei, Anhui 230026, China}

\author{Romain Couvreur}
\email{romain.couvreur@ens.fr}
\affiliation{Laboratoire de Physique Th\'eorique, D\'epartement de Physique de l'ENS, \'Ecole Normale Sup\'erieure, Sorbonne Universit\'e, CNRS, PSL Research University, 75005 Paris, France}
\affiliation{Sorbonne Universit\'e, \'Ecole Normale Sup\'erieure, CNRS, Laboratoire de Physique Th\'eorique (LPT ENS), 75005 Paris, France} 
\affiliation{Institut de Physique Theorique, CEA Saclay, 91191 Gif-sur-Yvette, France}

\author{Youjin Deng}
\email{yjdeng@ustc.edu.cn}
\affiliation{Hefei National Laboratory for Physical Sciences at Microscale and Department of Modern Physics, University of Science and Technology of China, Hefei, Anhui 230026, China}
\affiliation{CAS Center for Excellence and Synergetic Innovation Center in Quantum Information and Quantum Physics, University of Science and Technology of China, Hefei, Anhui 230026, China}

\author{Jesper Lykke Jacobsen}
\email{jesper.jacobsen@ens.fr}
\affiliation{Laboratoire de Physique Th\'eorique, D\'epartement de Physique de l'ENS, \'Ecole Normale Sup\'erieure, Sorbonne Universit\'e, CNRS, PSL Research University, 75005 Paris, France}
\affiliation{Sorbonne Universit\'e, \'Ecole Normale Sup\'erieure, CNRS, Laboratoire de Physique Th\'eorique (LPT ENS), 75005 Paris, France} 
\affiliation{Institut de Physique Theorique, CEA Saclay, 91191 Gif-sur-Yvette, France}

\date{\today}

\begin{abstract}

Percolation, a paradigmatic geometric system in various branches of physical sciences, is known to possess logarithmic factors in its correlators. Starting from its definition, as the $Q\rightarrow1$ limit of the $Q$-state Potts model with $S_Q$ symmetry, in terms of geometrical clusters, its operator content as $N$-cluster observables has been classified. We extensively simulate critical bond percolation in two and three dimensions and determine with high precision the $N$-cluster exponents and non-scalar features up to $N  \! =\! 4$ (2D) and $N  \! =\! 3$ (3D). The results are in excellent agreement with the predicted exact values in 2D, while such families of critical exponents have not been reported in 3D, to our knowledge. Finally, we demonstrate the validity of predictions about the logarithmic structure between the energy and two-cluster operators in 3D.
\end{abstract}
\maketitle

Statistical systems at criticality are scale invariant and usually characterized by the power-law decay of
 their correlation functions. In $d=2$ dimensions (2D),
Conformal Field Theory (CFT) succeeded in computing the corresponding critical exponents 
(and many finer details) of such systems. But scale invariance
is also compatible with {\em logarithmic} factors in the correlators \cite{Gurarie93}, 
which may appear if the theory is non-unitary and the scaling dimensions of two or more distinct
operators coincide. Non-unitarity may be due to: quenched disorder (e.g., in disordered electron gases realizing the integer quantum Hall
plateau transition), non-positive definite Boltzmann weights (at Lee-Yang singularities in hard lattice gases), or non-local observables (in percolation
and polymer models) \cite{GurarieLudwig,LCFTreview}. In the corresponding Logarithmic CFT (LCFT), the dilatation operator is non-diagonalizable and modes of the Noether current
(stress tensor) realize indecomposable representations of the symmetry algebra.

The logarithmic factors can be understood by treating the LCFT as a limit of usual CFTs. This approach is particularly powerful when the system has
additional discrete symmetries (apart from conformal invariance), whose irreducibles characterize the different operators that mix indecomposably in
the LCFT limit where their scaling dimensions collide. These ideas were developed by Cardy \cite{Cardy99} and first applied by him to disordered systems
that arise in the $N\to0$ limit of the replica approach (the symmetry being the $\mathcal{S}_N$ replica group). 
Another case of interest is the $Q$-state Potts model
(with $\mathcal{S}_Q$ symmetry)~\cite{Wu1982,Cardy99,Vasseur15}. For $Q\in\mathbb{N}$ its continuum limit is unitary, 
but non-unitarity results from extending the
definition to $Q\in\mathbb{R}$, via a high-temperature expansion, and defining non-local observables in terms of 
cluster connectivities \cite{FK}. 
Taking then $Q \to Q_0$, with $Q_0$ integer, leads to LCFTs, the case $Q_0 = 1$ being (bond) percolation. 
Refs.~\cite{VJS,VJ} classified the $\mathcal{S}_Q$ irreducible operators,
related them to cluster observables and unravelled the corresponding LCFT contents.
We stress that these works and this Letter apply to {\em bulk} LCFT, which is more challenging \cite{gl21,puzzle,JS18} than its boundary counterpart \cite{VJS11,Gori2017}.
Remarkably, in this context, some of the most salient structural properties (like: for each $Q_0$,
which operator mixings produce logarithmic factors, and in which correlators) turn out to be {\em independent} of dimension $d$.
For $d>2$, this route seems the only known semi-rigorous way of deriving exact results on the logarithmic structure of LCFTs.
The set of predictions was greatly enhanced in \cite{CJV} by extending the initial treatment \cite{VJS,VJ} 
of scalar operators (i.e., that transform trivially under
rotations) to include also non-scalar operators that realize general representations of $S_Q$.

The study of CFTs in $d>2$ has recently been thrusted into the limelight by the conformal bootstrap program. 
Using constraints of unitarity, this approach has
led to new insights and improved the precision of critical exponents for local operators in the 3D Ising model \cite{BootstrapIsing}. Non-unitary extensions have given access to a few dimensions
in 3D (e.g., in the Yang-Lee model \cite{GR} or percolation \cite{LeClair}), but sometimes on condition that an exponent was determined independently, either perturbatively \cite{LeClair} or numerically. The LCFT approach to $d>2$ is complementary to the bootstrap in many respects: it focuses on
non-local operators rather than local ones, and targets exact structural properties rather than bounds on the numerical values of exponents. To complete its
predictions, one needs in particular to determine the scaling dimensions of the non-local operators by an independent means.

In this Letter, we present a high-precision numerical study of non-local and logarithmic correlators in 2D and 3D percolation. 
Their logarithmic structure is found to agree with theoretical predictions \cite{CJV}, for any $d$. The same is true for the 2D critical exponents.
We determine the exponents and universal indecomposability parameters in 3D, and also investigate the non-trivial behavior of correlators under
rotations.

\paragraph{Percolation and the $Q \to 1$ Potts model.}
The partition function $Z$ of the $Q$-state Potts model with
interactions $-K\delta_{\sigma_i,\sigma_j}$ along edges, $(ij) \in E$,
can be rewritten as the random cluster model \cite{FK,Wu1982}
\begin{eqnarray}
Z=\sum_{A\subseteq E}Q^{k(A)}v^{|A|}.
\label{partitionFK}
\end{eqnarray}
Here $v={\rm e}^K-1$, $|A|$ is the number of edges in the subset $A$, and $k(A)$ is
the number of connected components (clusters) in the graph obtained by
deleting lattice edges not in $A$. Note that (\ref{partitionFK}) makes sense for
$Q \in \mathbb{R}$ and gives access to non-local correlators of
cluster connectivities. The limit $Q \to 1$ describes bond percolation.

The original interactions suppose $Q \in \mathbb{N}$ and have an $\mathcal{S}_Q$ symmetry.
The classification of $\mathcal{S}_Q$ irreducible observables can nonetheless be analytically continued to
$Q \in \mathbb{R}$, and leads to exact results on cluster correlators \cite{VJS,VJ,CJV}.
Below we study the probabilities that $N$ clusters propagate from one neighborhood to another at the percolation
threshold. We mainly use the square (cubic) lattice in 2D (3D).
Their percolation thresholds are $p_c (2 {\rm D})=1/2$ \cite{kesten1980} 
and $p_c (3 {\rm D})=0.248 \, 811 \, 85 (10) $ \cite{Lorenz1998,Wang2013,xu2014}. 

\paragraph{Observables.}

Let $ {\mathcal V}_i \equiv (i_1, i_2, \ldots, i_N)$ denote $N$ mutually disconnected lattice sites in a small neighborhood.
We usually take their positions to be aligned,
$\bfr_{i_{m+1}} \! = \! \bfr_{i_m} + \boldsymbol\delta$, with $m \! = \! 1,2,\ldots,N-1$. 
For $|\boldsymbol\delta|=1$ they are nearest neighbors.
Let another site set ${\mathcal V}_j \equiv (j_1, j_2, \ldots, j_N)$ be distant from ${\mathcal V}_i$ 
by $\bfr = \bfr_j-\bfr_i$, with $r=|\bfr| \gg 1$. 
We consider configurations in which $N$ distinct percolation clusters propagate from  ${\mathcal V}_i $  to  ${\mathcal V}_j $,
i.e., each cluster connects a site in ${\mathcal V}_i $ to another site in ${\mathcal V}_j $.
There are $N!$ such configurations, symbolically represented as  $\PPa$ and $\PPb$ for $N=2$, 
$\PPPa$, $\PPPb$, $\PPPc$, $\PPPd$, $\PPPe$ and $\PPPf$ for $N=3$, etc.

Appropriate linear combinations of the corresponding probabilities ($\bbp_{\PPa}$, $\bbp_{\PPb}$, etc.)
give access to the operator content of the underlying field theory \cite{VJS,CJV}. 
More precisely, these combinations correspond, in the continuum limit,
to the two-point function of an operator. This correspondence relies on the local ${\cal S}_{N}$ symmetry between the $N$ spins of ${\mathcal V}_i$
(or ${\mathcal V}_j$), and the ${\cal S}_Q$ of the Potts model. Note that ${\cal S}_Q$ is subtly non-trivial, since percolation is not $Q=1$ but rather $Q \to 1$.
The definitions of observables acting on $N=2$ and $N=3$ spins are recalled below. Each of them corresponds, technically, to
a pair of Young diagrams for ${\cal S}_{N}$ and ${\cal S}_Q$ \cite{CJV}.

Consider first observables describing the propagation of $N=2$ clusters. There are two different combinations,
corresponding to the \underline{s}ymmetric and \underline{a}ntisymmetric Young diagrams of ${\cal S}_2$,
\begin{eqnarray}
\bbp_{2{\rm s}} & =& \bbp_{\PPa} + \bbp_{\PPb}   \nonumber \\
\bbp_{2{\rm a}} & =& \bbp_{\PPa} - \bbp_{\PPb}  \; ,  \nonumber
\label{eq:N2}
\end{eqnarray}
corresponding in the continuum limit to the two-point functions of two operators $\mathcal{O}_{2{\rm s}}$ and $\mathcal{O}_{2{\rm a}}$.
Below, we also use the term observable to describe a two-point function. The scaling dimensions of these operators in 2D CFT are known \cite{CJV}.
Notice that $\mathcal{O}_{2{\rm s}}$ ($\mathcal{O}_{2{\rm a}}$) transforms trivially (non-trivially) under rotations: 
in 2D the latter has non-zero conformal spin,
$h-\bar{h} = 1$ (see (\ref{eq:Kac})).

For $N=3$ clusters, the relevant combinations are
\begin{eqnarray}
\bbp_{3{\rm s}}   & =& \np \bbp_{\PPPa} + \bbp_{\PPPb} + \bbp_{\PPPc} + \bbp_{\PPPd} + \bbp_{\PPPe} +\np \bbp_{\PPPf} \nonumber \\
\bbp_{3{\rm m}}  & =& 2 \bbp_{\PPPa} + \bbp_{\PPPb} + \bbp_{\PPPc} - \bbp_{\PPPd} - \bbp_{\PPPe} - 2 \bbp_{\PPPf} \nonumber \\
\bbp_{3{\rm a}}   & =& \np \bbp_{\PPPa} - \bbp_{\PPPb} - \bbp_{\PPPc} + \bbp_{\PPPd} + \bbp_{\PPPe} - \np \bbp_{\PPPf} \nonumber  \; ,
\label{eq:N3}
\end{eqnarray}
where $\bbp_{N\circ}$ (with subscript $\circ={\rm s}, {\rm m}, {\rm a}$) refers to the \underline{s}ymmetric, \underline{m}ixed and
\underline{a}ntisymmetric Young diagram of ${\cal S}_3$.
For $N=4$, we have $\bbp_{4 \rm s}$, $\bbp_{4 \rm m1}$, $\bbp_{4 \rm m2}$, $\bbp_{4 \rm m3}$ and $\bbp_{4 \rm a}$, 
since ${\cal S}_4$ admits
five Young diagrams; see the Supplementary Material (SM) for details. 
All these observables, qua two-point functions, are expected to decay algebraically at criticality, as $r^{-2 \Delta}$,
with (a priori) distinct, symmetry-dependent scaling dimensions.

\paragraph{Critical exponents in 2D.}

In 2D, the exponents can be computed exactly using algebraic methods and CFT results. They are expressed in terms of conformal weights
$h_{r,s}$ in the so-called Kac parameterization
\begin{eqnarray}
h_{r,s}=\frac{(r(x+1)-s x)^2-1}{4x(x+1)} \;,
\label{eq:Kac}
\end{eqnarray}
where $x$ determines $Q$ by $\sqrt{Q}=2\cos\frac{\pi}{x+1}$ (so $x=2$ for percolation), and $(r,s)$ are Kac labels.

The exponents of the above observables were already studied in \cite{CJV}, but in the geometry of an infinite cylinder, suitable for transfer matrix (TM) computations.
In that case, ${\cal V}_i$ (${\cal V}_j$) reside at the lower (upper) rim of the cylinder. Since moreover clusters cannot cross, certain configurations
cannot be realized on the cylinder. In this Letter we perform Monte Carlo (MC) computations in physically more relevant geometry of the plane.
This alleviates these restrictions, leading in some cases to changes in the exponents.

On the cylinder, the symmetry ${\cal S}_N$ is effectively restricted to its subgroup of cyclic permutations ${\cal C}_N$.
The scaling dimension related to the one-dimensional representation $\exp(i2\pi p/N)$ of ${\cal C}_N$ was found to be \cite{CJV}
\begin{eqnarray}
\Delta_{p,N}=h_{p/N,N}+h_{-p/N,N} \;,
\label{DimensionJTL}
\end{eqnarray}
where $p$ is an integer between $\lfloor-N/2\rfloor$ and $\lfloor N/2\rfloor$ determined as follows:
For the Young diagram of a given operator, find all its corresponding standard Young tableaux and compute for each of them its index ${\cal I}$ (= sum of descents).
The $p$ in (\ref{DimensionJTL}) is then the value of ${\cal I} \, \mbox{mod} \, N$ leading to the smallest $\Delta_{p,N}$ (most relevant contribution).

\begin{figure}
\includegraphics[width=\linewidth]{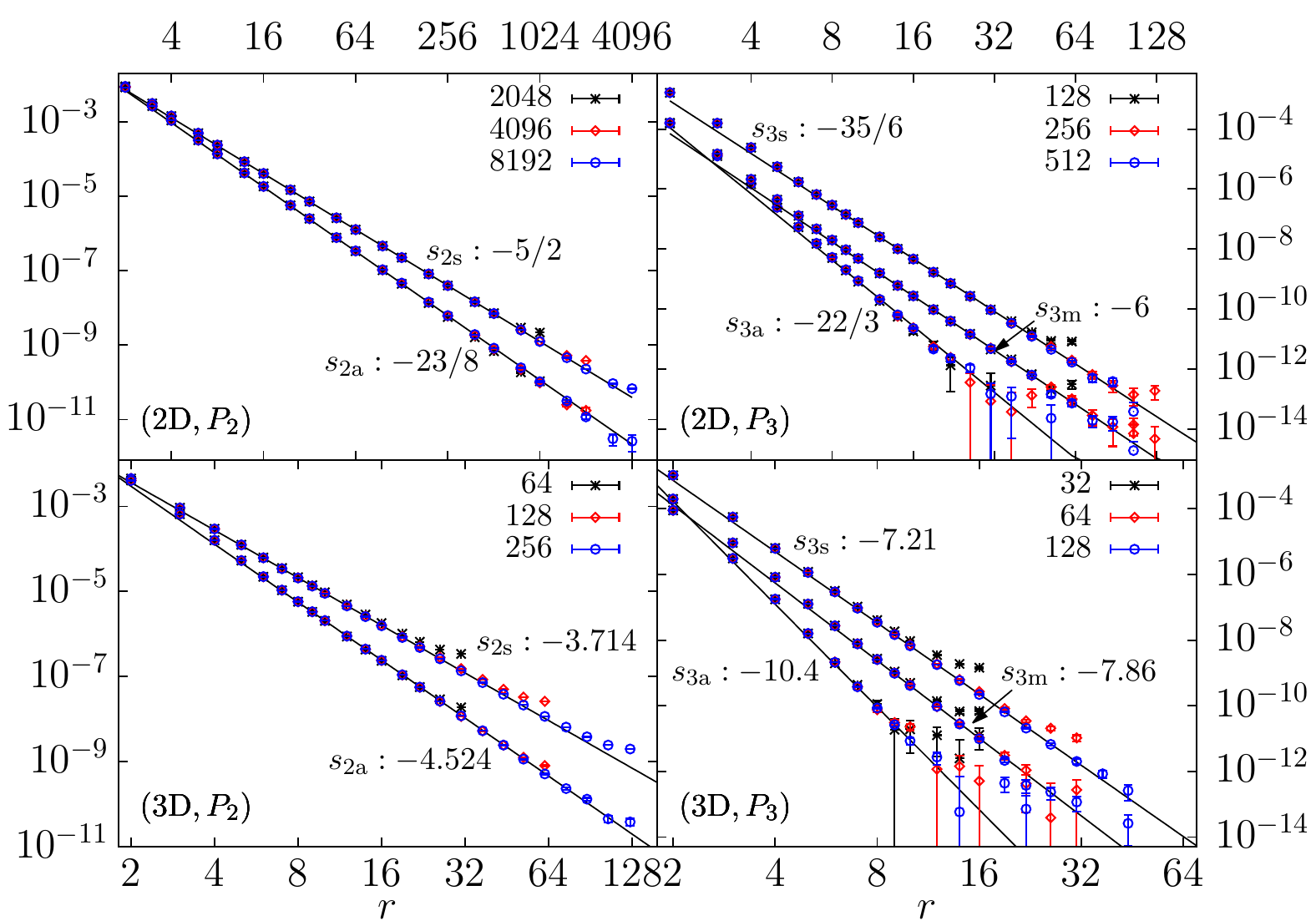}
\caption{Log-log plot of $\bbp_{2 \circ}$  and $\bbp_{3 \circ}$ versus distance $r$ in 2D and 3D,
for different system sizes $L$.  Subscripts $\circ={\rm s}, {\rm m}, {\rm a}$ refer to
\underline{s}ymmetric, \underline{m}ixed and \underline{a}ntisymmetric correlators.
Straight lines with slope $s$ come from the least-squares fits.
For clarity,  $\bbp_{\rm 3{\rm s}}$ ($\bbp_{\rm 3{\rm m}}$) have been multiplied by a factor 5 (2).
}
\label{fig01}
\end{figure}

Consider first the $N=2$ observables. The scaling dimensions of $\mathcal{O}_{\rm 2s}$ and $\mathcal{O}_{\rm 2a}$ are
$\Delta_{\rm 2s}=2h_{0,2}=5/4$ and $\Delta_{\rm 2a}=h_{1/2,2}+h_{-1/2,2}=23/16$. The leading behavior of the probabilities are then
$\bbp_{2{\rm s}} \propto r^{-2\Delta_{\rm 2s}}$ and $\quad\bbp_{2{\rm a}} \propto r^{-2\Delta_{\rm 2a}}$. 
These predictions were checked by TM on the cylinder \cite{CJV},
and they are confirmed by our MC computations in the plane (upper-left panel of Fig.~\eqref{fig01} and Tab.~\ref{tab:exponent}).

The $N=3$ case is more interesting, since the restriction from ${\cal S}_3$ to ${\cal C}_3$ does not necessarily hold in the geometry relevant for MC.
In particular, both of $\PPPa$ and $\PPPf$ can generically be realized in the plane, while on the cylinder one of them cannot. 
To be in the generic situation, the points in ${\cal V}_i$ and ${\cal V}_j$ must be sufficiently spaced,
and we henceforth assume this is the case. In this case, the restriction from ${\cal S}_N$ to ${\cal C}_N$---a key argument in
deriving (\ref{DimensionJTL})---does not occur. But remarkably, (\ref{DimensionJTL}) still appears to provide
the correct scaling dimension, provided $p$ is chosen differently (see below).

The case of the symmetric operator $\mathcal{O}_{3{\rm s}}$ presents no such subtleties.
Its corresponding Young diagram has one index ${\cal I} = 0$, and setting $p=0$ in (\ref{DimensionJTL}) we find $\Delta_{3{\rm s}}=2h_{0,3}=35/12$.
This coincides with the well-known six-leg watermelon operator \cite{SD87} and agrees well with the MC results for $\bbp_{3{\rm s}}$
(upper-right panel of Fig.~\eqref{fig01} and and Tab.~\ref{tab:exponent}).
Similarly, for $\mathcal{O}_{3{\rm m}}$ we find $p=1$, and $\Delta_{3{\rm m}}=h_{1/3,3}+h_{-1/3,3}=3$
agrees with the numerics for $\bbp_{3{\rm m}}$. The interesting case concerns $\mathcal{O}_{3{\rm a}}$, for which ${\cal I}=3$. 
On the cylinder, one finds the {\em same} scaling dimension as for $\mathcal{O}_{3{\rm s}}$, namely $\Delta_{0,3} = 35/12$,
since $p = {\cal I} \, \mbox{mod} \, N = 0$. However, our MC results in the plane unambiguously agree with $\Delta_{3,3} = 11/3$.
We hypothesize that exact results in the plane are obtained by setting $p = {\cal I}$ (without mod $N$). This is confirmed by
 an exhaustive study of the five $N=4$ exponents (see SM and Tab.~\ref{tab:exponent}).

\paragraph{Critical exponents in 3D.}

The definitions of the observables are independent of $d$. The corresponding operators are only quasi-primary in 3D,
but the various probabilities $\bbp$ should still scale with distinct scaling dimensions, due to their different symmetry content.
This is confirmed by our MC results (bottom panels of Fig.~\eqref{fig01} and Tab.~\ref{tab:exponent}).
Non-local operators of this type do not appear to have been previously studied in 3D.

\begin{table}[htbp]
\caption{\label{tab:exponent} Least-squares fitting results for $N$-cluster exponents $\Delta$ in 2D and 3D. The rows ``Theo." 
    are for the exact values from the $d=2$ LCFT. }
\begin{ruledtabular}
\begin{tabular}{llllll}
 2D      & $\Delta_{\rm 2s}$       & $\Delta_{\rm 2a}$      & $\Delta_{\rm 3s}$        & $\Delta_{\rm 3m}$     & $\Delta_{\rm 3a}$      \\
            & $1.2503(6)$        & $1.438 (4)$         & $2.93(4)$             & $2.986(14)$        & $3.75(20)$            \\ 
 Theo.   & $5/4$                  & $23/16$               & $35/12$                & $3$                      & $11/3$              \\
 \hline 
 2D      & $\Delta_{\rm 4s}$      & $\Delta_{\rm 4m1}$     & $\Delta_{\rm 4m2}$    & $\Delta_{\rm 4m3}$     & $\Delta_{\rm 4a}$           \\
           & $5.24(3)$           & $5.25(10)$              & $5.40(10)$            & $5.60(20)$           & $7.00(30)$           \\
    Theo.   &  21/4       & 339/64     & 87/16   & 363/64   & 111/16          \\
 \hline
 3D      & $\Delta_{\rm 2s}$     & $\Delta_{\rm 2a}$          & $\Delta_{\rm 3s}$     & $\Delta_{\rm 3m}$        & $\Delta_{\rm 3a}$ \\
            & $1.857(2)$        & $2.262(10)$            & $3.605(8)$        & $3.93(4)$               & $5.2(2)$ 
\end{tabular}
\end{ruledtabular}
\end{table}

\paragraph{Conformal spin.}

Observables that are not fully symmetric transform non-trivially under local rotations.
We first illustrate this in 2D, where we can compute the conformal spin of operators.
The spin corresponding to \eqref{DimensionJTL} is $|h_{p/N,N}-h_{-p/N,N}|=p$.
Thus, the $N=2$ operators $\mathcal{O}_{2s}$ and $\mathcal{O}_{2a}$ have spin $0$ and $1$, respectively.

\begin{figure}
\includegraphics[width=\linewidth]{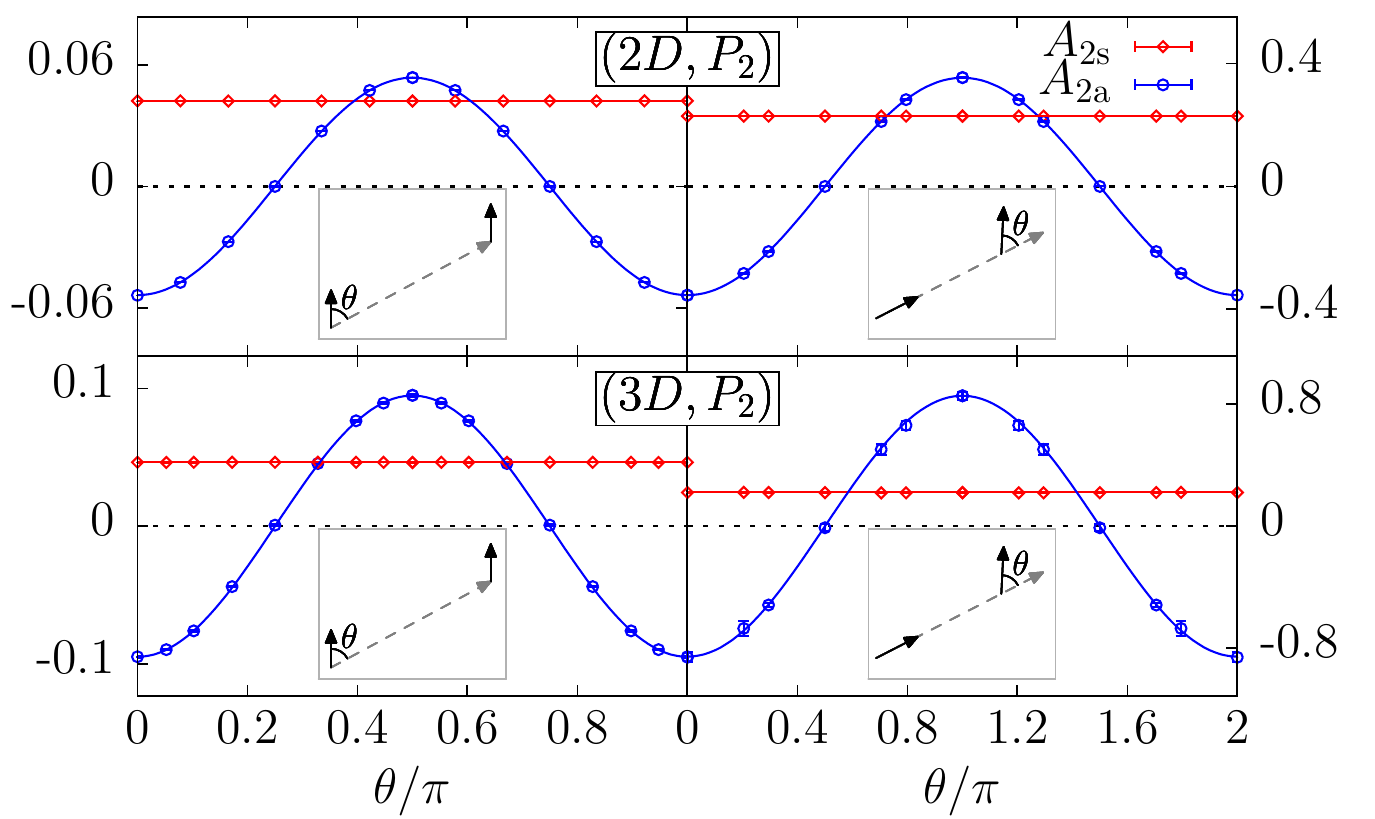}
\caption{Rotational dependence of $\bbp_{\rm 2s} r^{2\Delta_{\rm 2s}}$ and $\bbp_{\rm 2a} r^{2\Delta_{\rm 2a}}$. 
The results are obtained from extrapolation first to $L \rightarrow \infty$ and then to $r \rightarrow \infty$.
For both ways of rotations shown in the insets, the symmetric correlation $ \bbp_{\rm 2s} $ is independent of rotation angle $\theta$, 
    while the asymmetric one $\bbp_{\rm 2a}$ is proportional to $\cos (2\theta+\pi)$ or $\cos(\theta+\pi)$.
}
\label{fig03}
\end{figure}

This can be checked in MC simulations, using techniques similar to \cite{CJV}.
We perform a rotation of ${\cal V}_i$ around ${\cal V}_j$, while keeping the local orientation of each site in the neighborhoods fixed.
The upper-left panel of Fig.~\ref{fig03} shows the renormalized amplitude of the observables as a function of the angle.
It is clear that in the large-$r$ limit the two-point function of $\mathcal{O}_{2s}$ is invariant under rotations, whereas the
two-point function of $\mathcal{O}_{2a}$ fits perfectly with $\propto \cos(2\theta +\pi)$.
Another way to investigate the spin is to keep ${\cal V}_i$ around ${\cal V}_j$ fixed, but do a local
rotation of the relative position $\boldsymbol\delta$ in one of them. In CFT, this corresponds to a
simple conformal change of the metric around one point. This cannot be done continuously on the lattice,
but nevertheless the results are clear and match now $\propto \cos(\theta+\pi)$ for the spin-$1$ case (top-right panel Fig.~\ref{fig03}).
The factor-two difference of the periods is understood (see \cite{CJV} for details).

In 3D, we cannot compute the conformal spin from (\ref{DimensionJTL}), but on general grounds we expect it to be independent of $d$.
This is confirmed by the bottom panels of Fig.~\ref{fig03} that show the same two protocols as above, but now in 3D.

\begin{figure}
\includegraphics[width=\linewidth]{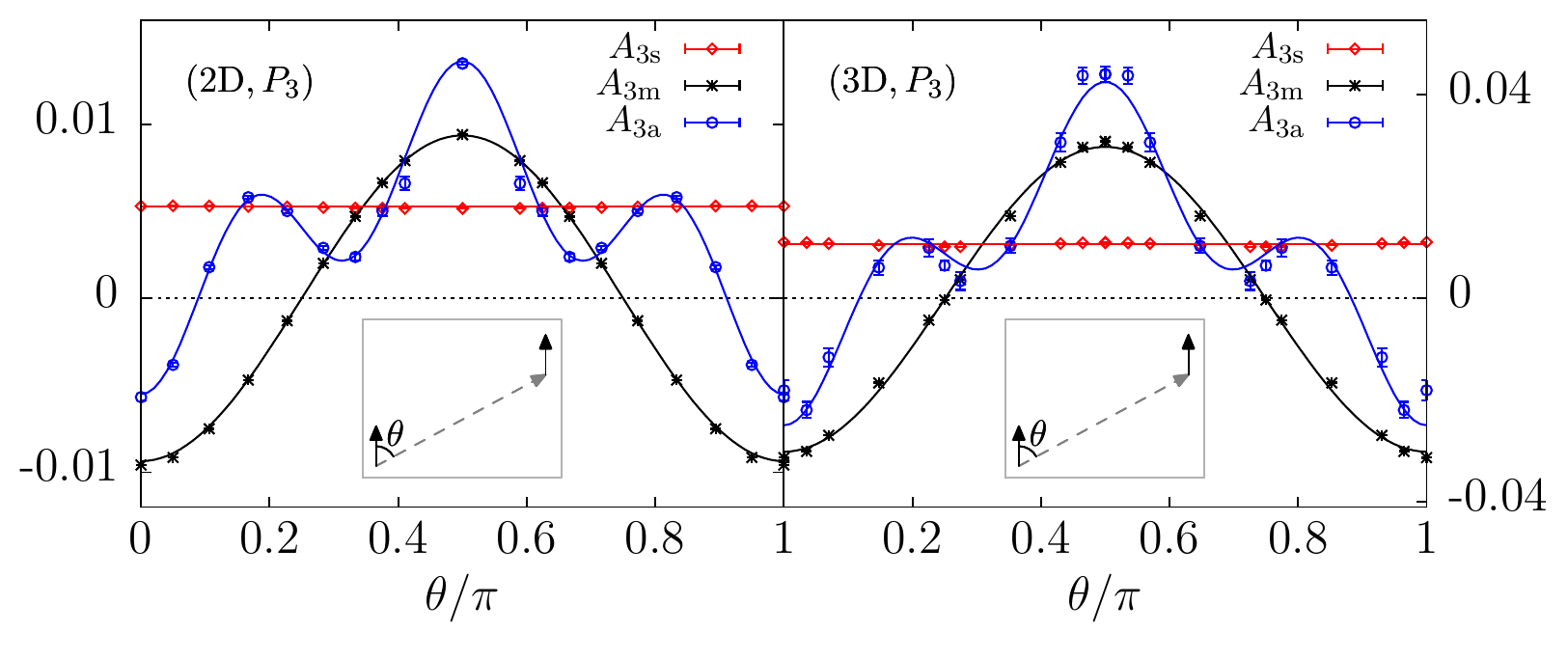}
\caption{Rotational dependence of $\bP_{\rm 3s}r^{2\Delta_{\rm 3s}}$, $\bP_{\rm 3 m}r^{2\Delta_{\rm 3m}}$
and $\bP_{\rm 3a}r^{2\Delta_{\rm 3a}}$. 
The rotation schemes are given in the insets.
The simulation parameters are $(L,r)=(8192,12)$ in 2D and $(512,9)$ in 3D.
To have a better view, the  $\bP_{\rm 3 m}$ data are rescaled  by a factor of $0.22$ in 3D.}
\label {fig04}
\end{figure}

The $N=3$ observables can be investigated in the same way. In 2D, the prediction is that $\mathcal{O}_{3{\rm s}}$ has spin 0,
while $\mathcal{O}_{3{\rm m}}$ has spin $1$, and $\mathcal{O}_{3{\rm a}}$ has spin $3$ in the plane and spin $0$ on the cylinder.
For numerical and practical reasons, we only investigated the first of the above protocols (the second being much more challenging
for $N=3$, given the constraints of the square lattice).  The results are shown in Fig.~\ref{fig04}. The spins of $\mathcal{O}_{3{\rm s}}$ 
    and $\mathcal{O}_{3{\rm m}}$ are seen to be 0 and 1, respectively (we find $\bbp_{3{\rm m}}\propto \cos(2\theta+\pi)$), in agreement with the
2D CFT predictions. The prediction for $\mathcal{O}_{3{\rm a}}$ is spin 3, but instead of a ``pure'' function of the form $\cos(2s\theta)$ with $s=3$,
$\bbp_{3{\rm a}}$ is seen to be a mixed sum
\begin{eqnarray} 
\bP_{\rm 3a} & \propto & \cos(2  \theta)+b_{1} \cos(4 \theta) + b_{3} \cos(6 \theta) \; , \nonumber 
\end{eqnarray} 
where $b_1$ and $b_2$ are non-universal amplitudes. In general, we expect the angular dependence of the two-point function of a spin-$s$ operator
to be a sum of the form $\sum_{k=1}^sb_k\cos(2k\theta)$, where the $b_k$ are non-universal. 
For all three observables, we obtain the same spin in 3D as in 2D, as expected.

\paragraph{Logarithmic features.}

A important breakthrough of \cite{VJS} was to prove the possibility of studying LCFTs through a limiting procedure.
In this class of theories---which includes percolation---there exist operators whose two-point functions are not purely a power-law.
In 2D, they have a logarithmic dependence of the form
\begin{eqnarray}
\left<\mathcal{O}_\Delta(0)\mathcal{O}_\Delta(r)\right>=\frac{\theta-2b\log r}{r^{2\Delta}} \;,
\label{logscaling}
\end{eqnarray}
where $\Delta$ is the scaling dimension, $\theta$ is non-universal, and $b$ is called an indecomposability parameter.
We know that $b$ is universal, and exact results are known for its value in many cases \cite{GurarieLudwig,VJS11,puzzle,gl21}.

For a LCFT to result as a limit of ordinary CFTs, the scaling dimensions of two operators must collide in the limit.
This is exactly what happens for 2D percolation in two dimensions: the scaling dimensions of the local energy operator $\varepsilon$
and of the symmetric two-cluster operator $\mathcal{O}_{2{\rm s}}$ collide when $Q \to 1$. This is accompanied by
a divergence in the two-point function of $\mathcal{O}_{2{\rm s}}$, which can be removed by mixing the two operators
into a Jordan cell. The parameter $b$ is proportional to the quantity
\begin{eqnarray}
 \delta =2\times\lim_{Q\rightarrow1}\frac{\Delta_2-\Delta_{\varepsilon}}{Q-1} \;.
\label{formulaIndec}
\end{eqnarray}
This universal number characterizes both the LCFT at $Q=1$ and the limit of CFTs when $Q\to 1$.

It is possible to go further and isolate the logarithmic factor in (\ref{logscaling}).
    For $N=2$, let $\bP_0 \equiv \bP_{\PPzero}$ be the probability each of the four specified points belongs to a different percolation cluster;
let $ \bP_1$ be the probability that the points belong to three different clusters, one of which propagates from one site in ${\mathcal V}_i$
    to another site in $ {\mathcal V}_j$, viz.\ $ \bP_1 \equiv \bP_{\PPca} + \bP_{\PPcb}+\bP_{\PPcc}+\bP_{\PPcd} $.
    Note that $\bP_{\PPzero}$ increases with $r$ and converges to $(\bP_{\neq})^2$ for $r \rightarrow \infty$, where $\bP_{\neq}$ is the
probability that the two points in ${\cal V}_i$ belong to different percolation clusters. 
The main result of \cite{VJS} is that the composite observable
\begin{equation} 
F(r)=\frac{  \bP_0(r) + \bP_1(r) - (\bP_{\neq})^2}{\bP_{2{\rm s}}(r)} \sim \delta \log (r)\; ,
\label{purelogscaling}
\end{equation}
diverges as a pure logarithm. Crucially, the predictions (\ref{formulaIndec})--(\ref{purelogscaling}), as well as
$\Delta_2 = \Delta_{\varepsilon}$ exactly at $Q=1$, hold in both 2D and 3D. In 2D we know also
$\Delta_2(Q)$ and $\Delta_{\varepsilon}(Q)$,
\begin{eqnarray}
\Delta_{\varepsilon}=2h_{2,1},\quad\Delta_2=2h_{0,2}
\end{eqnarray}
so $\delta = 2 \sqrt{3}/\pi\approx1.10266\ldots$.

\begin{figure}
\begin{center}
\includegraphics[width=1.0\linewidth]{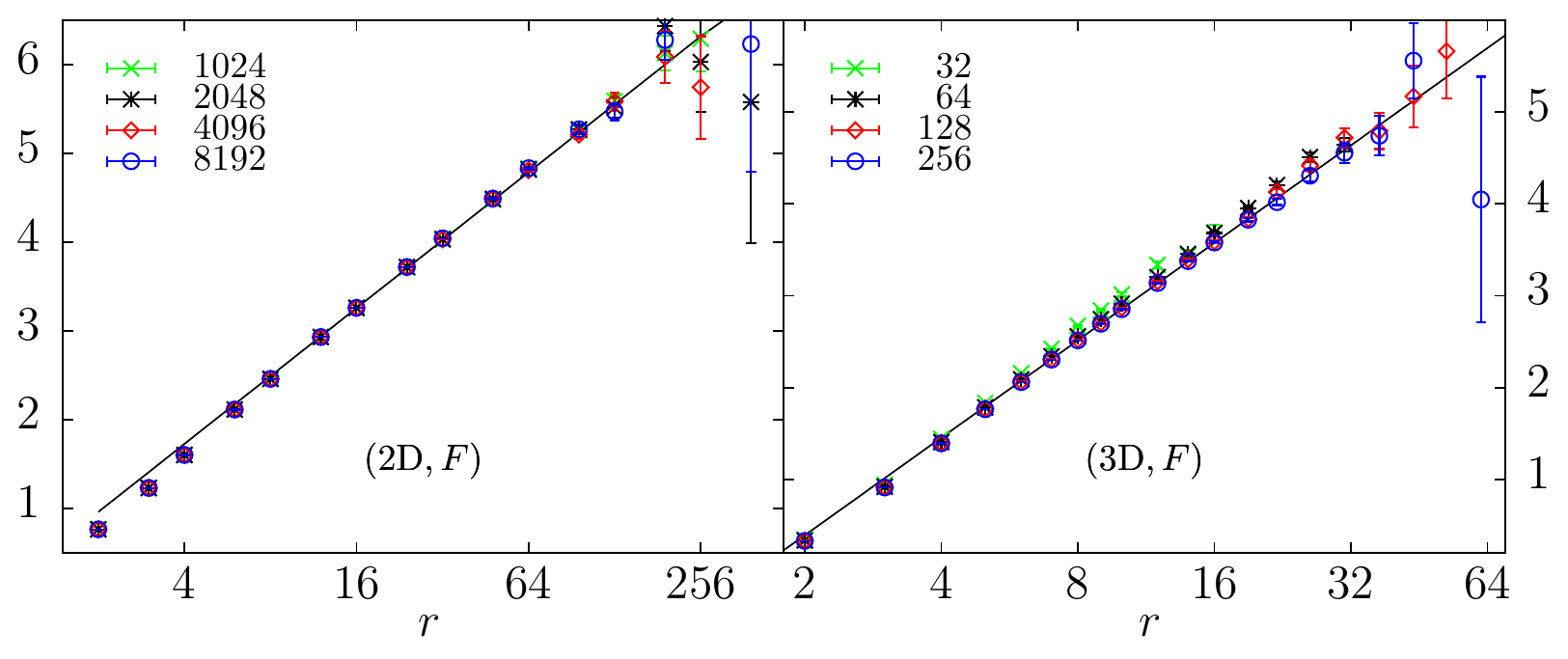}
    \caption{Semi-log plot of the logarithmic correlation $F(r)$ for 2D and 3D. 
    The slopes of the straight lines are universal, and the values are respectively $2\sqrt{3}/\pi$ and $1.53(3)$.
    }
\label {fig02}
\end{center}
\end{figure}

In the numerics we take each of $ {\mathcal V}_i$ and $ {\mathcal V}_j$ 
to contain a pair of nearest-neighbor sites. 
From the least-squares fit of the data in Fig.~\ref{fig02}, we obtain $\delta(2{\rm D}) = $ 1.12(3), improving the
numerics in \cite{VJS}, and very close to the exact result.
The 3D logarithmic scaling in (\ref{purelogscaling}) is
confirmed very clearly in the right panel of Fig.~\ref{fig02}, and we find $\delta(3{\rm D}) = $ 1.53(3).
Of course, $\Delta_2(Q)$ and $\Delta_{\varepsilon}(Q)$ are not known analytically in 3D.
We can nonetheless give a rough estimate of $\delta$ from (\ref{formulaIndec}) by using the numerical values
$\Delta_2 \approx 2.243(2)$ and $\Delta_\varepsilon \approx 1.413(1)$ for the 3D Ising model ($Q=2$)~\cite{Deng2004,Ferrenberg2018}.
This gives $\delta \approx 2(\Delta_2(Q=2)-\Delta_\varepsilon(Q=2))=1.66$. The agreement
with $\delta(3{\rm D})$
is surprisingly good, showing that $\Delta_2(Q)$ and $\Delta_{\varepsilon}(Q)$ have
little curvature (as in 2D).

\paragraph{Numerical details.}

Our Monte Carlo simulations are carried out for bond percolation on the square (2D) and cubic
(3D) lattices, with toroidal boundary conditions and linear system sizes varying from $L = 8$ to 8192 in 2D and 256 in 3D. 
It is especially challenging that the correlators 
decay very rapidly with $r$, in particular for $N=3$ and 4.
For instance, $\bP_{\rm 3{\rm a}}$ decays with exponent  $2\Delta_{\rm 3{\rm a}}({\rm 3D}) \approx 10.4$, 
so $\bP_{\rm 3{\rm a}} \leq10^{-14}$ already for $ r \approx 24$; see Fig.~\ref{fig01}.
Thus, reliable data are only available for a small range of $r$ and $L$, calling for careful
finite-size analysis.

As a first step, configurations of percolation clusters are produced by the standard procedure.
The whole lattice is visited site by site, starting from $i_1 \in {\cal V}_i$, 
and the $N$-cluster correlation functions are measured. 
Since most CPU time is spent measuring, independent simulations 
are performed for each $N$. In total, we used  $\approx 2 \times 10^6$ CPU core hours (see SM for further details).

Data for $N=2, 3$ (with $|\boldsymbol\delta| = 1$) are partly shown in Fig.~\ref{fig01}, and fitted to
$ \mathcal{O} (r)|_L= r^{-2 \Delta_{\mathcal{O}}} (a+b_1r^{-1} +b_2 r^{-2})$ by the least-squares criterion. 
For fixed $L$, we impose cutoffs, $r_{\rm min} \leq r \leq r_{\rm max}$,
on the data admitted in the fit, and we study the effect on the residual $\chi^2$ 
of varying $r_{\rm min}$ and $r_{\rm max}$. 
Results are then extrapolated to $L \rightarrow \infty$.
To avoid simultaneous finite-$r$ and finite-$L$ corrections,
we also simulate $r = \alpha L$, 
with $ 0 < \alpha < 1$ constant. Those data are fitted by $\mathcal{O} (L)
= (\alpha L)^{-2 \Delta_{\mathcal{O}}} (a+b_1 L ^{-1} +b_2 L^{-2})$.
Final results are reported in Tab.~\ref{tab:exponent}, where the quoted error bars include systematic uncertainties.

\paragraph{Discussion and outlook.} 
To summarize, we verified and completed exact predictions about cluster exponents for percolation in 2D. The dependence of certain exponents on the geometry (cylinder or plane) is noteworthy. In 3D we gave estimates for new exponents, and theoretical predictions \cite{VJS} of the logarithmic structure were ve\-ri\-fied. We find that the rotational behavior of correlators is similar in 3D and 2D, with spin $s=p$ in both cases.
While our analysis is confined to percolation, it proves the avail of studying LCFT as a limit of ordinary CFT. Being one of the few methods to study LCFT in higher dimensions, we believe it opens the possibility to derive exact results in 3D for a wide range of models.

\section*{ Acknowledgments.}  XJT and YD thank the support by National
Natural Science Foundation of China (grant 11625522) and the
Ministry of Science and Technology of China (grant 2016YFA0301604). 
RC and JLJ were supported by the ERC Advanced Grant NuQFT.

\bibliographystyle{apsrev4-1}
\bibliography{mybib}

\newpage
\clearpage
\onecolumngrid
\section{Supplemental Material}
\onecolumngrid
This Supplemental Material contains additional details about the results given in the main text. We first give the definitions of the $N=4$-cluster observables and detail the computation of the 2D critical exponents from the corresponding Young diagrams. The following section contains technical details about the simulations and the fitting scheme, including a few additional plots.
\section{\rom{1}. ADDITIONAL DEFINITIONS}
Using the same notation as in the main text, we define the $5$ observables corresponding to the $N=4$ correlation functions $P_{\rm 4\circ}(\{\rm \circ=s,m1,m2,m3,a\})$:

\begin{equation}
\begin{split}
    \bP_{\rm 4s} =&\;\; \bP_{\Pabcd}+  \np \bP_{\Pabdc} +  \np \bP_{\Pacbd} + \bP_{\Padbc} + \np \bP_{\Pacdb} +\np \bP_{\Padcb} +\np \bP_ {\Pbacd} +\np \bP_{\Pbadc}  +\np \bP_{\Pcabd}+ \bP_{\Pdabc} +\np \bP_{\Pcadb} + \np \bP_{\Pdacb} \nonumber \\
    +&\;\; \bP_{\Pbcad} +\np \bP_{\Pbdac} + \np \bP_{\Pcbad} + \bP_{\Pdbac} + \np \bP_{\Pcdab}+\np \bP_{\Pdcab} +\np \bP_{\Pbcda} + \np \bP_{\Pbdca} +\np\bP_{\Pcbda} + \bP_{\Pdbca} +\np \bP_{\Pcdba} +\np \bP_{\Pdcba}  \\ 
    \bP_{\rm 4m1} = & 3 \bP_{\Pabcd}- \np \bP_{\Pabdc} +3\bP_{\Pacbd} - \bP_{\Padbc} - \np \bP_{\Pacdb} -\np \bP_{\Padcb} +  3\bP_{\Pbacd} - \np \bP_{\Pbadc} +3\bP_{\Pcabd}- \bP_{\Pdabc} -\np \bP_{\Pcadb} - \np \bP_{\Pdacb}   \\ 
    +&  3\bP_{\Pbcad} - \np\bP_{\Pbdac}+ 3 \bP_{\Pcbad} - \bP_{\Pdbac} - \np \bP_{ \Pcdab}- \np \bP_{\Pdcab} -\np \bP_{\Pbcda} -\np \bP_{\Pbdca} -\np \bP_{\Pcbda} - \bP_{\Pdbca} -\np \bP_{\Pcdba} -\np \bP_{\Pdcba} \\ 
    \bP_{\rm 4m2} =& 2\bP_{\Pabcd} + 2\bP_{\Pabdc} -\np \bP_{ \Pacbd} - \bP_{\Padbc} -\np \bP_{\Pacdb} -\np\bP_{ \Padcb} +  2\bP_{\Pbacd} + 2\bP_{\Pbadc} -\np \bP_{ \Pcabd}- \bP_{\Pdabc} -\np \bP_{ \Pcadb} -\np \bP_{\Pdacb} \\
    - &\np \bP_{\Pbcad} - \np \bP_{ \Pbdac} - \np \bP_{ \Pcbad} - \bP_{\Pdbac} + 2\bP_{\Pcdab}+ 2\bP_{ \Pdcab} -\np \bP_{ \Pbcda} -\np \bP_{\Pbdca} -\np \bP_{\Pcbda} -\bP_{\Pdbca} + 2\bP_{\Pcdba} + 2\bP_{\Pdcba}\\ 
    \bP_{\rm 4m3} =& 2\bP_{\Pabcd}-2\bP_{\Pabdc} -\np \bP_{ \Pacbd} + \bP_{\Padbc} +\np \bP_{\Pacdb} -\np \bP_{\Padcb} +  2\bP_{\Pbacd} -2\bP_{\Pbadc} -\np \bP_{ \Pcabd}+ \bP_{\Pdabc} +\np \bP_{\Pcadb} -\np  \bP_{\Pdacb}  \\
    -&\np \bP_{\Pbcad} - \np \bP_{\Pbdac} + \np \bP_{\Pcbad} + \bP_{\Pdbac}  + 0\bP_{\Pcdab}+ 0\bP_{\Pdcab} +\np \bP_{\Pbcda} - \np \bP_{ \Pbdca} +\np \bP_{ \Pcbda} - \bP_{\Pdbca} +0 \bP_{\Pcdba} +0 \bP_{\Pdcba} \\ 
    \bP_{\rm 4a} = &\np \bP_{\Pabcd}- \np\bP_{ \Pabdc} -\np \bP_{ \Pacbd} + \bP_{\Padbc} +\np \bP_{ \Pacdb} - \np \bP_{ \Padcb} -\np \bP_{ \Pbacd} +\np \bP_{ \Pbadc}  +\np \bP_{ \Pcabd}- \bP_{\Pdabc} -\np \bP_{ \Pcadb} +\np \bP_{\Pdacb}  \\
    + &\np \bP_{\Pbcad} -\np \bP_{\Pbdac} -\np  \bP_{ \Pcbad} + \bP_{\Pdbac}  +\np \bP_{ \Pcdab}-\np \bP_{\Pdcab} -\np \bP_{ \Pbcda} + \np \bP_{ \Pbdca} +\np \bP_{  \Pcbda} - \bP_{\Pdbca} -\np \bP_{ \Pcdba} +\np \bP_{ \Pdcba } ,\\
\end{split}
\end{equation}

each corresponding to a Young diagram for a representation of ${\cal S}_4$.

The computation of critical exponents involves algebraic manipulations of Young tableaux. Therefore, before discussing each of the five observables, we recall a few definitions. A Young diagram with $N$ boxes is specified by listing its row lengths (in non-increasing order, and summing up to $N$) between square brackets. A standard Young tableau associated to such a Young diagram is obtained by filling the boxes with the integers from $1$ to $N$, such that the numbers are increasing in each row and column. The {\em index} of a standard Young tableau is defined as the sum of its {\em descents}, where a number $i$ is said to be a descent if $i+1$ appears in a row strictly below $i$. In the following, for clarity, descents in a drawing of a standard Young tableaux will be written in bold font. We now discuss each of the $5$ possibilities for the $N=4$-cluster observables. More details about the algebraic derivation of the conformal dimensions can be found in \cite{CJV}. Note that the conformal dimensions are given in the geometry of the plane. We thus do not restrict the symmetry to ${\cal C}_4$ (as on a cylinder) and follow the computational scheme described in the main text.

\begin{itemize}
 \item The first observable is $\bP_{\rm 4s}$ corresponding to the fully symmetric representation of ${\cal S}_4$. Its Young diagram is $[4]$ and its only standard tableau is 
\begin{equation}
\ytableausetup{boxsize=2em,centertableaux}
\begin{ytableau}
 1 & 2 & 3 & 4
\end{ytableau}
\end{equation}
with an index $\mathcal{I}=0$. The predicted dimension in 2D is then $\Delta_{0,4}=21/4$ with a spin $0$.

 \item The observable $\bP_{\rm 4m1}$ corresponds to the Young diagram $[3,1]$. It has $3$ distinct standard Young tableaux, and the one corresponding to the lowest index (hence the lowest conformal dimension) is
\begin{equation}
\ytableausetup{boxsize=2em,centertableaux}
\begin{ytableau}
 {\bf 1} & 3 & 4 \\
 2
\end{ytableau}
\end{equation}
with an index $\mathcal{I}=1$. The predicted dimension in 2D is $\Delta_{1,4}=339/64$ with a spin $1$.

\item The observable $\bP_{\rm 4m2}$ corresponds to the Young diagram $[2,2]$. It has $2$ distinct standard Young tableaux and the one corresponding to the lowest index is
\begin{equation}
\ytableausetup{boxsize=2em,centertableaux}
\begin{ytableau}
 1 & {\bf 2}  \\
 3 & 4
\end{ytableau}
\end{equation}
with an index $\mathcal{I}=2$. The predicted dimension in 2D is $\Delta_{2,4}=87/16$ with a spin $2$.

\item The observable $\bP_{\rm 4m3}$ corresponds to the Young diagram $[2,1,1]$. It has $3$ distinct standard Young tableaux and the one corresponding to the lowest index is
\begin{equation}
\ytableausetup{boxsize=2em,centertableaux}
\begin{ytableau}
 {\bf 1} & 4  \\
 {\bf 2} \\
 3
\end{ytableau}
\end{equation}
with an index $\mathcal{I}=3$. The predicted dimension in 2D is $\Delta_{3,4}=363/64$ with a spin $3$.

\item The last observable is $\bP_{\rm 4a}$ corresponding to the Young diagram $[1,1,1,1]$. It has $1$ standard Young tableau
\begin{equation}
\ytableausetup{boxsize=2em,centertableaux}
\begin{ytableau}
 {\bf 1}  \\
 {\bf 2} \\
 {\bf 3} \\
 4
\end{ytableau}
\end{equation}
with an index $\mathcal{I}=6$. The predicted dimension in 2D is $\Delta_{6,4}=111/16$ with a spin $6$.
\end{itemize}
We provide numerical data supporting these results below.

\section{\rom{2}. Simulation details}
{\bf The $N=2$ and 3 cases.} The simulations for $N=2$ and 3  were carried out for the bond percolation 
on the square and the simple-cubic lattice at their percolation thresholds, $p_c (\rm 2D) =1/2$~\cite{kesten1980}
and $p_c(\rm 3D) = 0.248 \, 811 \, 85(10)$~\cite{Wang2013,xu2014} respectively.  Periodic boundary conditions were applied, 
and the linear system size was taken  in the range $ L \in [8, 8192]$ for 2D and $ [4, 512]$ for 3D. 
The $N$ lattice sites in the neighborhoods $\mathcal{V}_i$ and $\mathcal{V}_j$ were set to be aligned, separated by 
a unit lattice spacing ($|\boldsymbol\delta|=1$). The displacement $\mathbf{r}$ between $\mathcal{V}_i$ and $\mathcal{V}_j$ 
was taken perpendicular to $\boldsymbol{\delta}$, corresponding to $\theta=\pi/2$ in Figs.~2 (Left panel) and Fig.~3 in the main text. 

For a given linear size $L$ we sampled the $N$-cluster connectivities as functions of distance 
$r = |\mathbf{r}|$ and then defined the corresponding $N$-cluster two-point correlation functions $\bbp_{N}(r_k)$,
with  distance $r_k \in [2,L/2]$ for the $k$th data point. The increment of distance was set as $\delta r_k \equiv r_{k+1}-r_k = \max[1, 0.2r_k]$.
The log-log plot in Fig.~1 of the main tex confirms well that, both for 2D and 3D, the different correlation functions $\bbp_{N\circ}$ 
are governed by distinct scaling dimensions $\Delta_{N\circ}$, with $\circ=$ s, m, or a.
The up-bending deviation for large $r$ (close to $L/2$) is due to the periodic boundary conditions. 
We fitted the Monte Carlo data for each $L$ to $ \mathcal{O} (r)|_L= r^{-2 \Delta_{\mathcal{O}}(L)} (a+b_1r^{-1} +b_2 r^{-2})$ by 
the least-squares criterion, with various cutoffs $r_{\rm min} \leq r \leq r_{\rm max}$. 
Then, the $\Delta_{\mathcal{O}}(L)$ values were extrapolated to the thermodynamic limit $L \rightarrow \infty $;
for most cases, such finite-size dependencies were found to be very weak. 

\begin{figure}[h]
\resizebox{12cm}{!}{
\includegraphics[width=\linewidth]{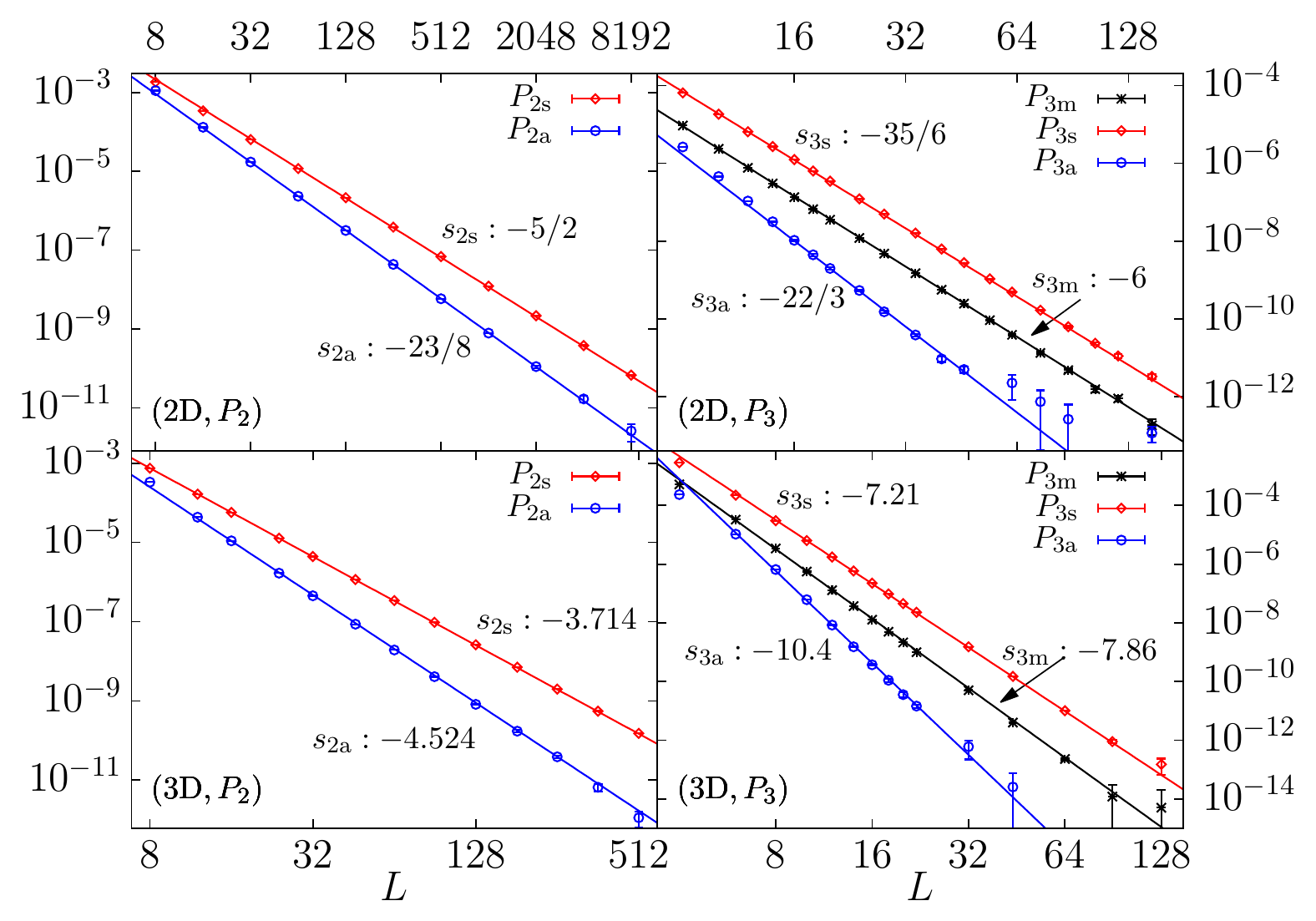}
}
\caption{\label{fig:alphaL} Log-log plot of $\bP_{\rm 2\circ}$  and $\bP_{\rm 3\circ}$ versus $L$ in 2D and 3D,
measured at $r=L/2$.
Subscripts $\circ={\rm s}, {\rm m}, {\rm a}$ refer to the
\underline{s}ymmetric, \underline{m}ixed and \underline{a}ntisymmetric correlators.
Straight lines with slope $s$ come from the least-squares fits.
For clarity,  $\bbp_{\rm 3s}$ ($\bbp_{\rm 3 m}$) has been multiplied by a factor 10 (2). 
}
\end{figure}

To avoid corrections stemming from both small $r$ and finite $L$, we carried out additional extensive simulations
and measured the correlation functions right at $r= L/2$. 
The standard scaling hypothesis leads to the finite-size scaling $ \mathcal{O} (L) \propto L^{-2 \Delta_{\mathcal{O}} }$.
The data in Fig.~\ref{fig:alphaL} were fitted to 
the ansatz  $\mathcal{O} (L) = L^{-2 \Delta_{\mathcal{O}}} (a+b_1 L ^{-1} +b_2 L^{-2})$.
The results are shown in Tab.~I in the main text, 
where the error bars include both 
the statistical error coming directly from the fit and the systematic error coming from the variation among the results for different cutoffs.


\begin{figure}
\resizebox{10cm}{!}{
\includegraphics[width=\linewidth]{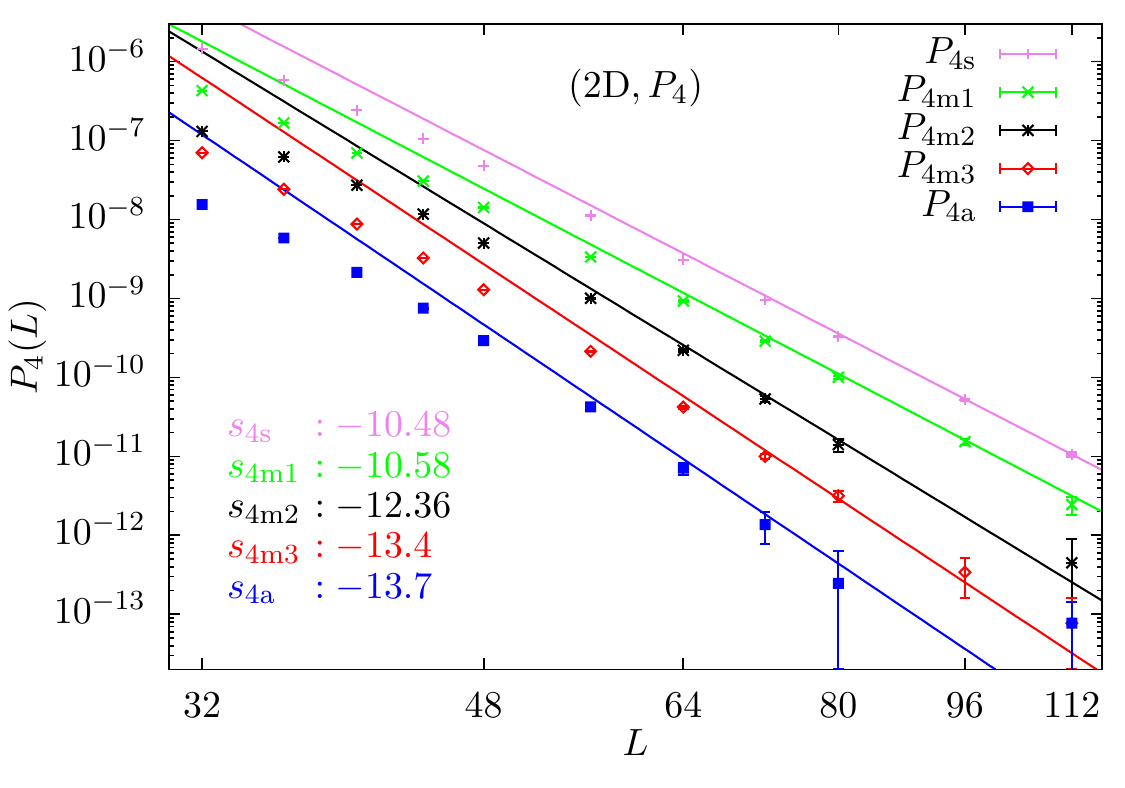}
}
\caption{\label{fig:4N} Log-log plot of $\bP_{\rm 4\circ}({\rm \circ=s,m1,m2,m3,a})$ versus distance $r$ for
critical bond percolation on the $L=2048$ triangular lattice.
The separation between neighboring sites in $\mathcal{V}_i$ and $\mathcal{V}_j$ is taken as $|\boldsymbol\delta| = 3\sqrt{3}$.
The values of the slopes $s$ for the straight lines are predicted from the Kac formula. 
For clarity,  $P_{\rm 4s}$ is multiplied by a factor of 3. 
    }
\end{figure}

{\bf The $N=4$ case.} The $N$-cluster correlation functions, particularly the asymmetric one $\bbp_{N {\rm a}}$, 
decay more and more rapidly when $N$ increases, and thus the simulations become more and more challenging. 
For $N=4$ we focused on the 2D case and measured $\bbp_{N {\rm o}}(r)$  for a fixed and sufficiently large system.
To control the corrections from both small $r$ and finite $L$, and to enhance the magnitude in front of the power-law decay, 
we performed some primary simulations on both the square and the triangular lattice, 
for different separations between neighboring sites in $\mathcal{V}_i$ (and $\mathcal{V}_j$).
We chose to simulate bond percolation on the triangular lattice at the percolation 
threshold $p_c=2\sin (\pi/18)$~\cite{Sykes}. 
The separation was taken as $|\boldsymbol\delta| = 3 \sqrt{3}$, and
the angle $\theta$ between $\boldsymbol\delta$ and $\mathbf{r}$ was still set at $\theta=\pi/2$.
About $1.2 \times 10^9$ independent percolation configurations were obtained for system size $L=2048$, 
and  $3 \times L^2$ measurements of $\bbp_{4 {\rm o}}$  were taken for each configuration. 
The $\bbp_{4 {\rm o}}$ data are shown in Fig.~\ref{fig:4N},  and the fitting results in Tab.~I of the main text
are consistent with the theoretical predictions $\Delta_{\rm 4s}=\Delta_{0,4}=21/4$, 
$\Delta_{\rm 4m1} =\Delta_{1,4}=339/64 \approx5.297$, $\Delta_{\rm 4m2}=\Delta_{2,4}= 87/16=5.4375$,
$\Delta_{\rm 4m3}=\Delta_{3,4}= 363/64 \approx 5.671$ and $\Delta_{\rm 4a}=\Delta_{6,4}=111/16\approx6.937$.

\begin{figure}
\resizebox{11cm}{!}{
\includegraphics[width=\linewidth]{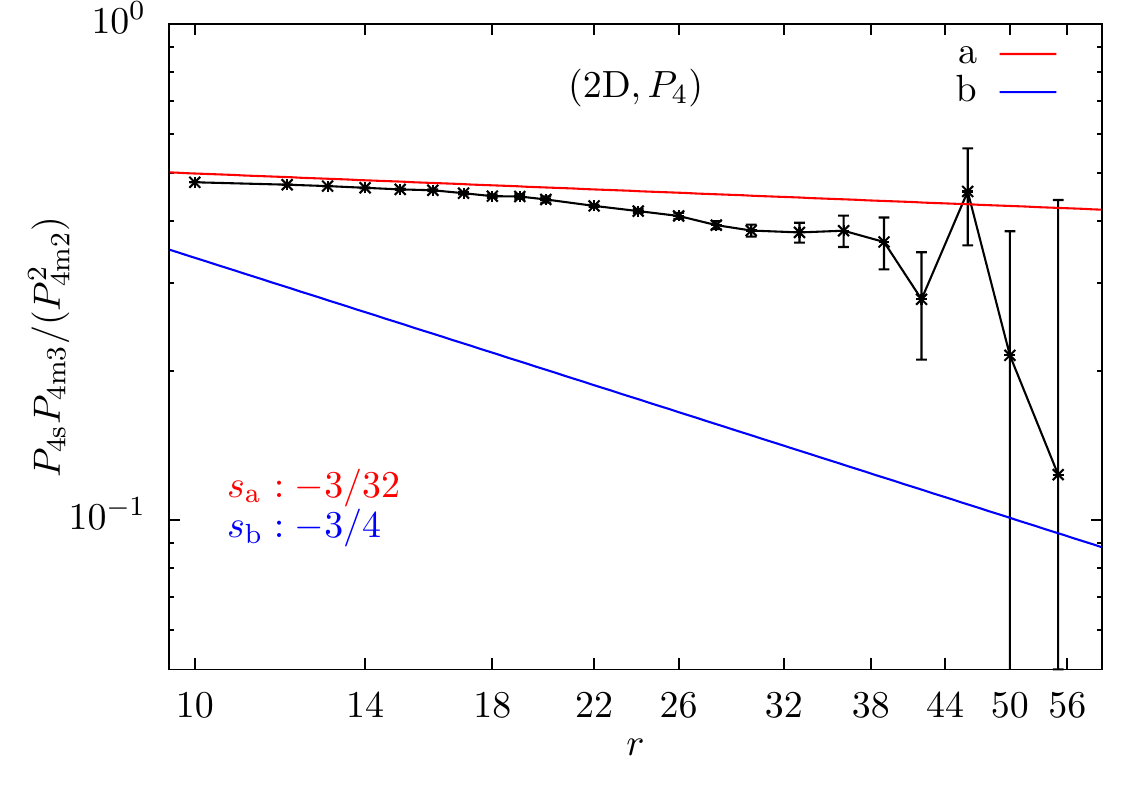}
}
\caption{
    \label{fig:4Nratio}
    Log-log plot of the ratio $(P_{\rm 4s} P_{\rm 4m3})/P_{\rm 4m2}^2$  versus distance $r$. 
    Assuming the theoretical values $\Delta_{\rm 4s}=21/4$, $\Delta_{\rm 4m2}=87/16$ and $\Delta_{\rm 4m3} = 363/64$,
    the slope should be $-3/32$ (red line). On the other hand, the slope would be $-3/4$ if $\Delta_{\rm 4m3} = 6$, 
    two error bars away from the fitting result $\Delta_{\rm 4m3} = 5.60(20)$. 
    The plot suggests that $\Delta_{\rm 4m3} = 6$ is much less likely than $\Delta_{\rm 4m3} = 363/64$.
    }
\end{figure}

We can further perform some self-consistency checks of the fitting results. 
For instance, we define the ratio $P_{\rm 4s} P_{\rm 4m3}/P_{\rm 4m2}^2$, in which the amplitude of corrections 
may be diminished due to partial cancellation between the numerator and the denominator. 
Assuming that the theoretical predictions hold exactly for $P_{\rm 4s}$, $P_{\rm 4m2}$ and $P_{\rm 4m3}$, 
the ratio would decay as $r^{-3/32}$, which is consistent with Fig.~\ref{fig:4Nratio}. 
On the other hand, if one assumes that $\Delta_{\rm 4m3} = \Delta_{5,4}=6$---a putative value that would be consistent with the general Kac formula,
but at a wrong value of the index, situated here two error bars away from our final numerical estimate $\Delta_{4m3}=5.60(20)$---the
log-log plot in Fig.~\ref{fig:4Nratio} would have a slope $-3/4$. This is seen to be rather clearly ruled out by the actual numerical data.


\end{document}